\documentclass{emulateapj}
\usepackage{graphics}
\usepackage{multirow}

\newcommand{\sech}{{\rm sech}}

\shorttitle{EFFECTS OF THE ISM MODEL IN GALACTIC DISCS}
\shortauthors{TASKER AND BRYAN}

\begin{document}

\title{The Effect of the Interstellar Model on Star Formation Properties in Galactic Disks}

\author{Elizabeth J. Tasker\altaffilmark{1,2}}
\author{Greg L. Bryan\altaffilmark{2}}

\altaffiltext{1}{Department of Astronomy, University of Florida, Gainesville, FL 32611}
\altaffiltext{2}{Department of Astronomy, Columbia University, New York, NY 10027}

\begin{abstract}
We studied the effect of interstellar gas conditions on global galaxy simulations by considering three different models for the ISM. Our first model included only radiative cooling down to $300$\,K, our second model added an additional background heating term due to photoelectric heating, and our third model uses an isothermal equation of state with a temperature of $10^4$\,K and no explicit heating or cooling. Two common prescriptions for star formation are implemented in each case. The first is based on cosmological simulations with a low threshold for star formation but also a low efficiency. The second assumes stars form only in high density regions but with a higher efficiency. We also explore the effects of including feedback from type II supernovae.  We find that the different ISM types produce marked differences in the structure of the disk and temperature phases present in the gas, although inclusion of feedback largely dominates these effects.  In particular, size of the star-forming clumps was increased both by background heating and by enforcing an isothermal ISM. We also looked at the one dimensional profiles and found that a lognormal PDF provides a good fit for all our simulations over several orders of magnitude in density. Overall, despite noticeable structural differences, the star formation properties in the disk are largely insensitive to ISM type and agree reasonably well with observations.
\end{abstract}

\section{Introduction}

Star formation is one of the most perplexing processes in the galaxy due to its immense complexity on the very small scales and its apparent simplicity on the very large. What is difficult to know is to what extent the small scale physics can be ignored when considering the global evolution of galaxies.

Observationally, we know there is a simple relationship between the surface density of gas in a disk galaxy and the surface density of star formation \citep{Kennicutt1998, Kennicutt1989, Schmidt1959}. This should imply that star formation is a straight forward, universal process depending on average properties of the gas over many kiloparsecs.  The (perhaps naive) picture is one of gravity acting to collapse the gas into giant molecular clouds out of which stars form, a process which is hindered by rotational shear, thermal pressure, turbulence, magnetic fields, cosmic ray pressure and energy injected from supernovae \citep{MacLow2004}.  Yet when we start looking at star formation on small, parsec, scales, we find the gas out of which stars form to be a turbulent, multiphase medium, strongly affected by local conditions and not at all indicative of a global law.  

So complex is this interstellar medium (ISM) that simulations which model it have been forced to consider only small sections of the galaxy to achieve the required resolution \citep[e.g.][]{Slyz2005, Joung2006, deAvillez2004}. Until recently, simulations which model the entire galaxy have been forced to simplify the structure of the ISM to an isothermal or fixed two- or three-phase body for the calculation to be at all feasible \citep[e.g.][]{Harfst2006, Li2005b, Robertson2004, Semelin2002}. Not that such assumptions about the structure of the ISM are groundless. Analytical calculations performed by \citet{McKee1977} put together the now traditional picture of a three-phase ISM, largely filled by hot gas from supernovae explosions. Others have suggested modifications \citep[e.g.][]{Cox2005, Norman1989} but this basic scenario is still with us.

Yet while both of these approaches have been highly informative about the evolution and structure of galaxies they each have disadvantages. The small box simulations are unable to model global properties such as star formation histories, disk structure and the Schmidt law, resulting in their main comparison points being restricted to galaxies where we are able to observe and measure the ISM. Global models, on the other hand, have been unable to compute the evolution of the interstellar medium and the effect of this simplification on the disk galaxy's own evolution is unknown. 

Recently, however, simulations have been developed that are bridging this gap. While still short of the resolution attained in kiloparsec-sized simulation boxes, these models do achieve the refinement needed to model a complete galaxy disk that includes a multiphase ISM. This paper continues our work began in \citet{Tasker2006} (hereafter TB06) which examined a three-dimensional isolated galactic disk using an adaptive-mesh refinement technique.  In that paper, we found that a multiphase medium with  a large variety of temperatures and densities, in good agreement with simulations performed in two-dimensions by \citet{Wada2001} and smaller scaled three-dimensional runs by \citet{Wada2002}. These results supported the small box simulations in implying that the traditional view of the interstellar medium as a three-phase structure in strict pressure equilibrium has some validity but is a significant over-simplification. 

What is less clear is how important this is. The existence of the Schmidt law might imply that the details of the ISM are not necessary to achieve accurate modeling in global galaxy simulations. And indeed, the Schmidt law and the observed star formation threshold in disk galaxies \citep{Kennicutt1989} has been successfully reproduced both in models with a multiphase ISM (TB06) and those with a fixed isothermal ISM \citep{Li2005a}. On the other hand, the observed global structure of the ISM is far from an uninterrupted pool of gas. In addition to its turbulent nature, observation of the ISM in the Large Magellanic Clouds show a complex series of HI filaments riddled with holes and shells \citep{Kim1998}. Two of our neighboring spiral galaxies, M31 and M33 show numerous holes 40\,pc to 1\,kpc wide \citep{Deul1990}, and our own Milky Way produces plumes of gas that rise off the disk's surface \citep{Otte2003}. Not only can this not be modeled without global multiphase ISMs, it seems impossible that this structure cannot play an integral part in the disk's evolution. Moreover, the role of stellar feedback remains something of a mystery in both isolated galaxy simulations and in cosmological runs where incorrect modeling is frequently cited as a possible cause of discrepancy with observational data \citep[][and references therein]{Tasker2006b}. Improved modeling of the ISM on a global scale could result in a fuller understanding of feedback, ultimately allowing more accurate feedback routines in large cosmological simulations where resolution of the individual galaxies is not yet possible. 

In this paper we will compare global models of isolated disk galaxies with three distinct ISM types. The first of these will include radiative cooling down to $300$\,K, the second type will contain radiative cooling and a background photoelectric heating source and our final type will have an isothermal ISM at a constant temperature of $10^4$\,K. For each of these models, we will test two common prescriptions for star formation and the effects of feedback from type II supernovae. The resulting structures are contrasted and the star formation properties of the disks compared with observations. 

For these simulations, we use a high-resolution adaptive mesh refinement (AMR) code which includes a full treatment of self-gravity of the gas rather than the fixed potential which is often used and a more sophisticated treatment of star formation and feedback. Our simulations concentrate on hydrodynamical effects, ignoring magnetic fields and cosmic ray pressure. 

In section 2, we describe our computational approach, including details of the code we are using and lay out the initial conditions for the problem. Sections 3 and 4 will focus on the disk structure and the properties of the ISM while section 5 will look at comparisons with observations. 

\section{Numerical Methods}

\subsection{The Code}
\label{sec:enzo}

The simulations were performed using the hydrodynamics adaptive mesh refinement (AMR) code, \emph{Enzo}, described in \citet{Bryan1997, Bryan1999, Norman1999, Bryan2001} and \citet{OShea2004}. The AMR technique is particularly strong in this work where the resolution of a complex multiphase medium is of paramount importance. The grid cells form natural boundaries which allows gas with a range of temperatures to coexist and evolve \citep{Slyz2005, Tasker2006}. Other codes using particle-based techniques frequently have an over-mixing problem, causing unphysical radiative losses unless algorithmic steps are taken \citep{Marri2003, Springel2003}. As a result, the majority of previous simulations have been unable to properly model a multiphase interstellar medium, making them unable to access its importance in star formation. 

For these simulations, we use a three-dimensional periodic box of side $1$\,h$^{-1}$\,Mpc. With a root grid of $128^3$ and 8 levels of refinement; our smallest cell size, and therefore maximum resolution, is approximately 50\,pc. For our higher resolution run, this was decreased further to 25\,pc.

To evolve the gas through time, \emph{Enzo} used a three-dimensional version of the ZEUS hydrodynamics algorithm \citep{Stone1992}. Radiative gas cooling followed the cooling curve of \citet{Sarazin1987} down to temperatures of $10^4$\,K and then rates from \citet{Rosen1995} down to $300$\,K. (The exception to this are for the simulations performed with an isothermal equation of state for the gas, where no cooling is allowed). This bottom temperature threshold is still above what would be found in the dense molecular clouds, but \citet{Rosen1995} argued that this crudely compensates for physical processes not modeled, such as magnetic fields, turbulence, and cosmic-ray pressure. This temperature range does take us to the upper limit of the cold neutral medium \citet{Wolfire2003}, allowing us to sample a realistic spread of phases in the gas.  

In addition to cooling, the gas can also be heated through supernovae feedback (described below) and photoelectric heating. Photoelectric heating, that is the photo emission of UV-irradiated dust grains, is thought to be the dominant factor in the formation of the cold and warm neutral mediums \citep{Wolfire1995}. If photoelectric heating is turned on, \emph{Enzo} includes the term $\Gamma_\textrm{pe}=5.1\times 10^{-26}$\,ergs$^{-1}$ to the energy equation for the gas in a scheme based on \citet{Joung2006}, but without the dependence on the height above the disk adopted in that work. The value of $\Gamma_\textrm{pe}$ is dependent on the incident radiation field and, as such, could be tied with the star formation rate in the gas. However, since we are considering Milky Way-sized galaxy disks, we follow \citet{Joung2006} and adopt a number consistent with the local interstellar value. Since it is likely that this value was higher at earlier times, we also perform a run with a heating source of $\Gamma_\textrm{hs} = 1.41\times 10^{-24}$ \,ergs$^{-1} = 30\Gamma_\textrm{pe}$, a value that balances the cooling rate at densities of $1.0$\,cm$^{-3}$ at our initial temperature of $10^4$\,K. However, the increase in heating makes a relatively small difference to the star formation properties of the disk, so this run is not included in our main analysis but added to the discussion only. Note that we do not include photoionization heating, which would require full radiative transfer to model properly. 

\emph{Enzo} will form a star particle in a grid cell if it fulfills the following criteria \citep{Cen1992, OShea2004}: (i) the baryon density in the grid cell exceeds a designated threshold density, (ii) the mass of gas in the cell exceeds the local Jeans mass, (iii) there is convergent flow into the cell (i.e. $\nabla \cdot v < 0$) and (iv) the cooling time of the gas in the cell is less than its dynamical time ($\tau_{\textrm{cool}}<\tau_{\textrm{dyn}}$), or the gas is at the minimum temperature allowed of 300\,K. We will also consider one set of galaxy disks which are modeled with a purely isothermal interstellar medium. In this case, the above criteria are relaxed to only include the first three rules, since the gas is unable to cool. 

If a grid cell meets the required criteria for star formation, gas is removed from the cell and a star particle will be formed with mass calculated by:
\begin{equation}
m_{*}=\epsilon\frac{\Delta t}{t_{\textrm{dyn}}}\rho_{\textrm{gas}}\Delta x^3
\label{eq:sfr}
\end{equation}
where $\epsilon$ is the star formation efficiency (more properly the efficiency per dynamical time), $\Delta t$ is the size of the time step, $t_{\textrm{dyn}}$ is the time for dynamical collapse and  $\rho_{\textrm{gas}}$ is the gas density. A final, purely computational, criteria is that, even if a grid cell fulfills all the above criterion, a star particle will still not be formed if its mass is less than a given minimum value of (for this work) $10^4$\,M$_\odot$. Since the reason for this clause is purely numerical (a large number of small stars would greatly slow down the simulation) an override exists that creates a star below the minimum mass with a probability equal to the ratio between the mass of the would-be star particle and the minimum star mass. If this occurs, the resulting star mass is that of the minimum star mass or 80\% of the mass in the cell, whichever is smaller. The star particle's formation is then spread out over roughly a dynamical time, to mimic the formation of stars in the giant molecular clouds who follow the same time scale. 

The star particles themselves are modeled as a collisionless system using the N-body method. They gravitationally interact with the gas by mapping their positions onto the grid via a cloud-in-cell technique to produce a discretized density field. The number of star particles formed varies greatly depending on the values chosen for the threshold density and star formation efficiency described above. For our `C'-type star formation method where these values are low (see section~\ref{sec:runs}), roughly five million particles were used by the end of the simulation. In our `D'-type algorithm, these values are higher and the number of star particles formed is reduced to around 40,000.

\emph{Enzo} can also include stellar feedback from type II supernovae, often suggested as the main driving force for self-regulated star formation. When this feedback option is used, as it is in about half our simulations, then $10^{-5}$  of the rest-mass energy of generated stars is added to the gas' thermal energy over a time period equal to $t_{\textrm{dyn}}$. This is equivalent to a supernova of $10^{51}$ erg for every $55$\,M$_\odot$ of stars formed. The energy is deposited into the gas over one dynamical time or $10$\,Myrs, whicheve is longer.  During this period the energy is injected into the cell closest to the particle's current location.

 For all simulations, we adopt a cosmological model of a $\Lambda$CDM universe with $\Omega_m = 0.3$, $\Omega_\Lambda = 0.7$ and H$_0 = 67$\,kms$^{-1}$\,Mpc$^{-1}$.

\subsection{The Initial Conditions}

The initial conditions for our disk are the same as the simulations performed in TB06, with an ideal isothermal gas disk of temperature of $10^4$\,K and $\gamma = 1.67$ whose density profile is given by
\begin{equation}
\rho(r,z) = \rho_0 e^{-r/r_0}\sech^2\left(\frac{z}{2z_0}\right),
\end{equation}
sitting in a static dark matter halo. The major difference between these initial conditions and the ones set out in TB06 is the gas mass in the disk which we choose to be $6\times10^{10}$\,M$_\odot$, six times higher than in TB06, bringing it into line with the estimated total disk mass of the Milky Way (we do not begin with any stars). We also change the disk dimensions slightly from TB06, selecting a scale radius $r_0=3.5$\,kpc and a scale height $z_0=325$\,pc. Together with the gas mass, these choices fix the value for $\rho_0= 0.6$\,M$_\odot$pc$^{-3}$. Dark matter is included as a static halo potential in the form described by \citet{Navarro1997}.  The disk is initially borderline stable, with a Toomre Q parameter (described in section~\ref{sec:toomre}) of 0.5 at the center of the disk rising to 10 at the edge, but quickly cools to fragment. This is discussed in more detail in TB06.

\subsection{Summary of Performed Runs}
\label{sec:runs}

\begin{table}
\caption{Overview of simulations performed}
\begin{tabular}{rccp{1.05cm}cp{1cm}cccl}
& & & min $\Delta x$ \centering{(pc)} & $\epsilon$ & $n_{\rm thresh}$ (cm$^{-3}$) & Fb & Iso & Heat\\
\hline
\multirow{4}{*}{ISM \#1} & \multirow{4}{*}{\scalebox{1}[1.9]{$\Big\{$}} & C & \centering{50} & 0.05 & \centering{0.02} & No & No & No\\
&& D      & \centering{50} & 0.5  & \centering{$10^3$} & No   & No   & No\\
&& CFDBCK & \centering{50} & 0.05 & \centering{0.02}   & Yes  & No   & No\\
&& DFDBCK & \centering{50} & 0.5  & \centering{$10^3$} & Yes  & No   & No\\
\multirow{4}{*}{ISM \#2} & \multirow{4}{*}{\scalebox{1}[2.3]{$\Big\{$}} & HC & \centering{50} & 0.05 & \centering{0.02} & No & No & Yes\\
&& HD      & \centering{50} & 0.5  & \centering{$10^3$} & No   & No   & Yes\\
&& HCFDBCK & \centering{50} & 0.05 & \centering{0.02}   & Yes  & No   & Yes\\
&& HDFDBCK & \centering{50} & 0.5  & \centering{$10^3$} & Yes  & No   & Yes\\
&& HDHIRES & \centering{25} & 0.5  & \centering{$10^3$} & No   & No   & Yes \\
\multirow{3}{*}{ISM \#3} & \multirow{3}{*}{\scalebox{1}[1.4]{$\Big\{$}} & IC & \centering{50} & 0.05 & \centering{0.02} & No & Yes & No\\
&& ID      & \centering{50} & 0.5  & \centering{$10^3$} & No   & Yes  & No\\
&& IDJEANS  & \centering{50} & 0.5  & \centering{$10^3$} & No   & Yes  & No\\
\end{tabular}
\label{table:runs}
\end{table}

Our simulated galaxy disks are divided into three categories depending on the nature of their ISMs. For the first four simulations listed in table~\ref{table:runs}, the gas is allowed to radiatively cool down to $300$\,K via the cooling curves described in section~\ref{sec:enzo}. This is the same set-up (although for a heavier disk) that we presented in TB06. The next five runs also allow radiative cooling along the same curve, but include an additional photoelectric heating term, as described in section~\ref{sec:enzo}. The last three simulations use an isothermal equation of state, a popular assumption in global disk models where it has been hard to resolve a multiphase medium. The temperature of all the gas in these runs is fixed at $10^4$\,K. 

For each of these three ISM models, we consider two different star formation routines. These were presented in TB06 and we keep the same notation of `C' and `D' type as used in that paper. C-type star formation has an efficiency appropriate to the galactic disk as a whole, and one typically used in large-scale cosmological simulations (5\%). Because this is a global average, we use a low density threshold, allowing stars to form in relatively low density regions (providing they meet the criteria outlined in section~\ref{sec:enzo}). Since this means we do not follow the formation of the densest clumps, we use a Schmidt-like law to model the star formation rate. D-type star formation, on the other hand, makes use of the resolution in the disk and confines star formation to the densest structures which, at our resolution, compare to the largest giant molecular clouds. Stars therefore form only where the density is high ($10^3$\,cm$^{-3}$) but with a much higher efficiency per dynamical time of 50\%. 

Simulations which include stellar feedback from type II supernovae are performed for both the disks with radiative cooling (CFDBCK and DFDBCK for C-type and D-type star formation respectively) and for disks with radiative cooling and background heating (HCFDBCK and HDFDBCK). 

In addition to these simulations, we also perform runs designed to test the robustness of our results. HDHIRES is the same as HD, but with a root grid which has twice the spatial resolution and so eight times the mass resolution. IDJEANS uses identical conditions to ID, but adds another refinement criteria, resolving a cell when the Jeans length drops below four cell widths (at least until we reach the maximum refinement level), as suggested by \citet{Truelove1997}. As mentioned in section~\ref{sec:enzo}, we also perform a run with a higher heating rate. This is discussed in the discussion section at the end of this paper. 

The run times for these simulations depended greatly on whether stellar feedback was included. Without feedback, a run typically took $\sim 40$ hours on 8 processors of an Opteron Beowulf cluster. When feedback was introduced, the run took $\sim 130$ hours on 64 processors of a Xeon cluster. 

\section{The Structure of the Disk}
\label{sec:discstructure}

\begin{figure*} 
\begin{center} 
\includegraphics[width=\textwidth]{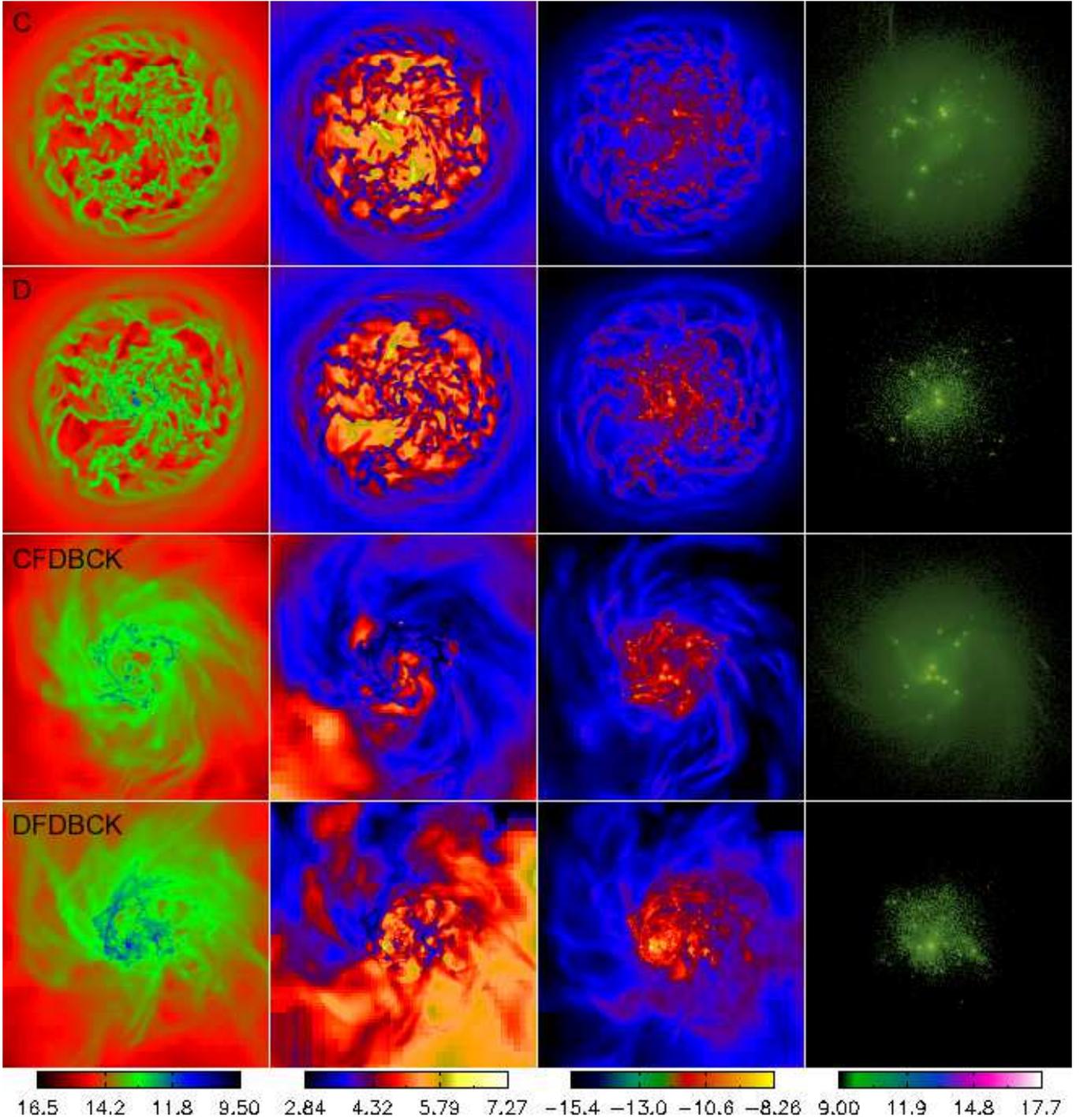}
\caption{Projections at 377\,Myrs of (left to right) gas density, temperature, pressure and stellar density for the runs with ISM~\#1 which include cooling but no photoelectric heating. Top to bottom, simulations shown are C, D, CFDBCK, DFDBCK. Images are 60\,kpc across. All scales are to the base-10 logarithm, and gas and star particle density is measured in M$_\odot$Mpc$^{-2}$, temperatures in K, and pressure on an arbitrary scale.\label{fig:projMW}}
\end{center} 
\end{figure*}

\begin{figure*} 
\begin{center} 
\includegraphics[width=\textwidth]{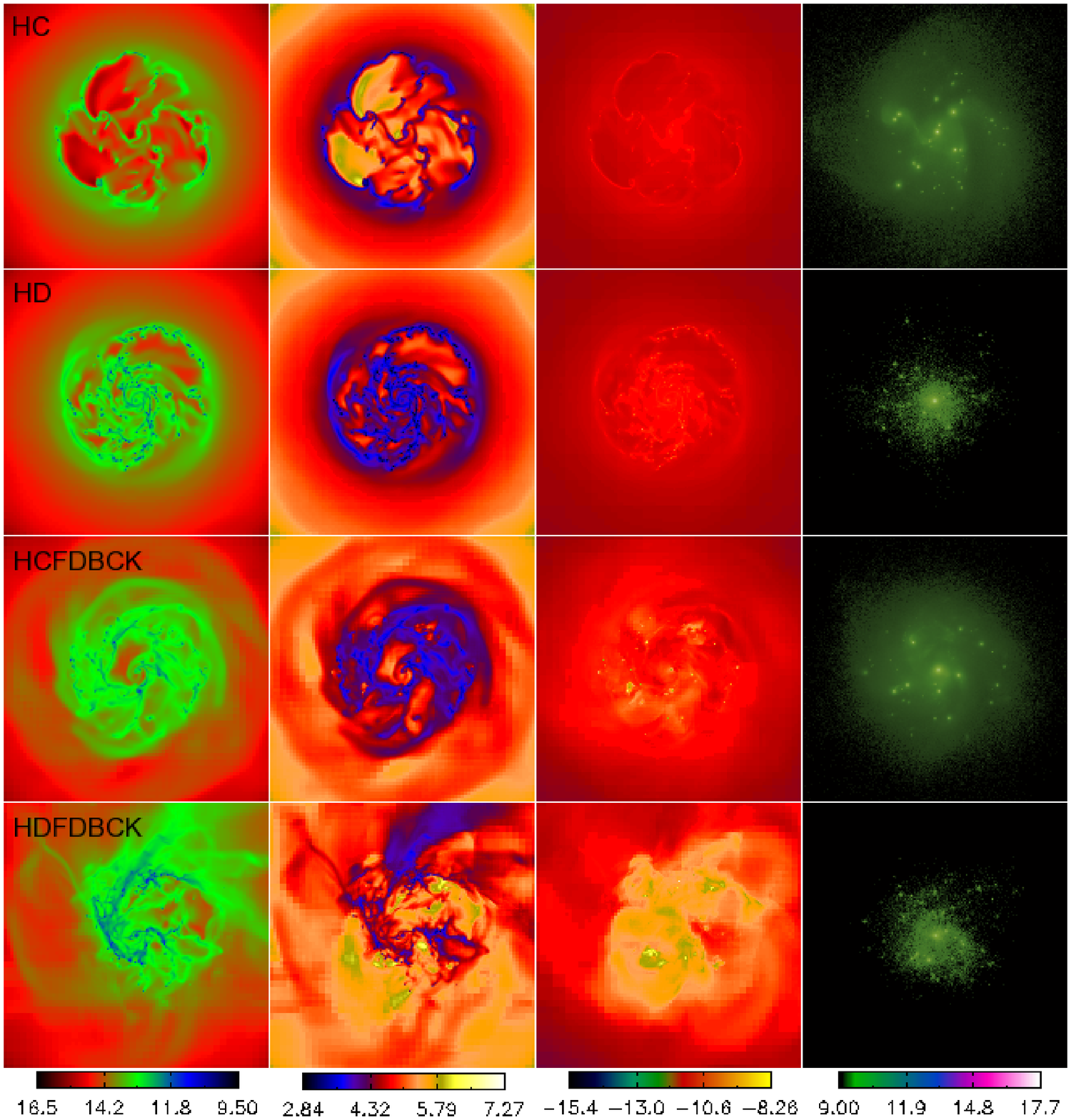}
\caption{Projections at 377\,Myrs of (left to right) gas density, temperature, pressure and stellar density for the runs with cooling and photoelectric heating. Top to bottom, simulations shown are HC, HD, HCFDBCK, HDFDBCK. Images are 60\,kpc across. All scales are to the base-10 logarithm, and gas and star particle density is measured in M$_odot$Mpc$^{-2}$, temperatures in K, and pressure on an arbitrary scale.
\label{fig:projHD}}
\end{center} 
\end{figure*}

\begin{figure*} 
\begin{center} 
\includegraphics[width=14cm]{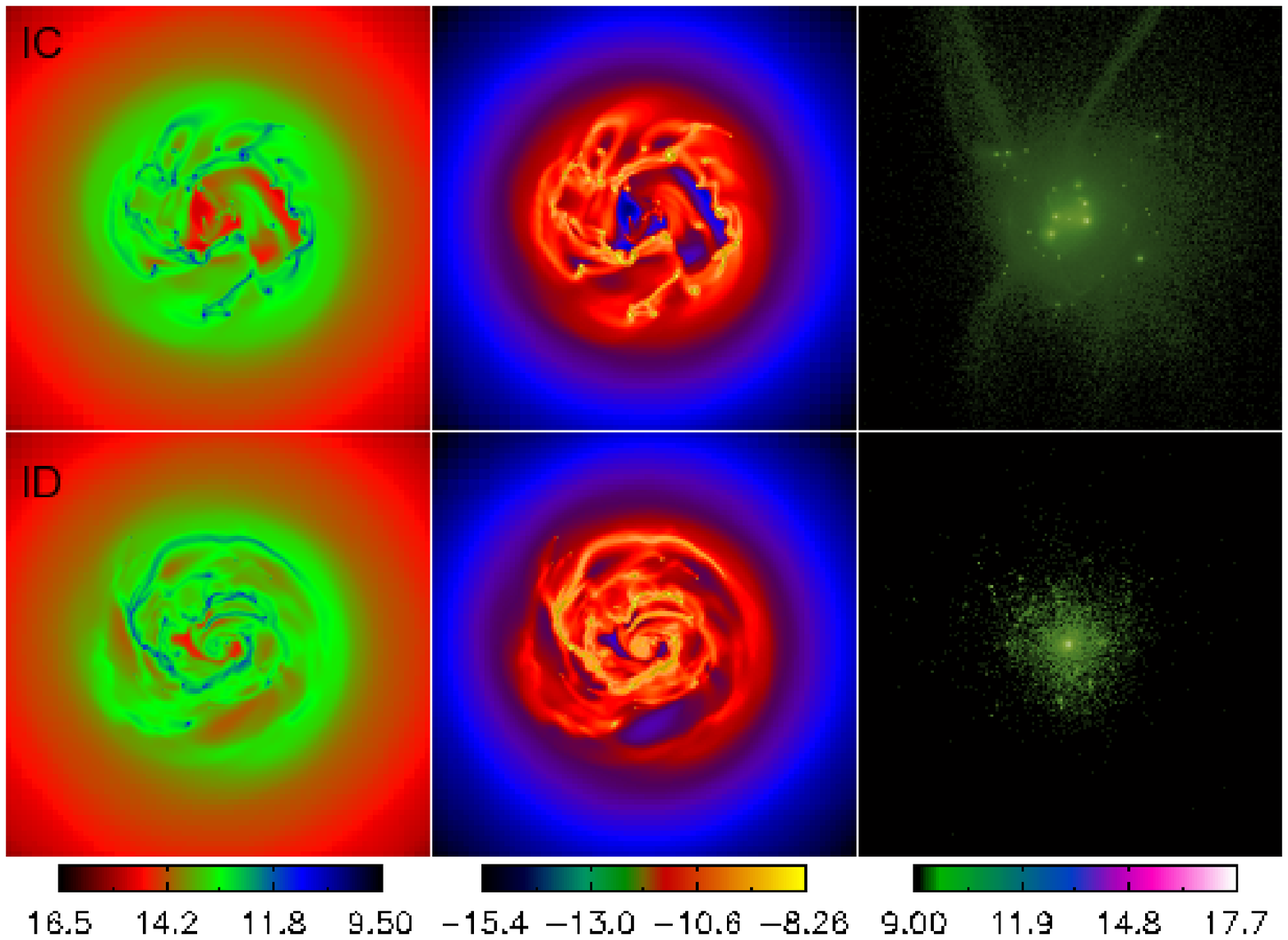}
\caption{Projections at 377\,Myrs of (left to right) gas density, pressure and stellar density for the runs with an isothermal ISM. Temperature is $10^4$\,K throughout simulation box. Simulations shown are IC (top) and ID. Images are 60\,kpc across. All scales are to the base-10 logarithm, and gas and star particle density is measured in M$_odot$Mpc$^{-2}$, temperatures in K, and pressure on an arbitrary scale.
\label{fig:projID}}
\end{center} 
\end{figure*}

During the initial evolution of our galaxy disks, we see the gas fragment, triggering a starburst. This process is described in detail in TB06, but for this work we want to avoid this primary collapse and compare our disks when they have reached a settled state but before any gas depletion becomes an issue. To judge where this point is, we examine the gas density in the disk over time. In the first 200\,Myrs, we see a clear and rapid drop in density corresponding to the initial fragmentation. The density then settles down, decreasing slowly as stars are formed until after around 500\,Myrs were the profile starts to deteriorate, especially in the non-feedback cases, as the majority of the gas is now converted to stars. Our images and plots are therefore largely taken between 200 - 400\,Myrs when the initial starburst is over and where the disks still have a substantial supply of gas. 

\subsection{Imaging the ISM}
\label{sec:images}

The global structure of the ISM can be viewed using face-on projections of the disk as shown in Figures~\ref{fig:projMW}, \ref{fig:projHD} and \ref{fig:projID}. Figure~\ref{fig:projMW} shows projections for our first ISM type, the top four simulations listed in table~\ref{table:runs}, all of which include radiative cooling but no background heating. From left to right, the images show gas density, temperature, pressure and stellar density. The top two columns which shows runs without feedback clearly show the region where the gas is gravitationally unstable. As in TB06, we see that the central gas has collapsed to form dense filaments and knots out to a clearly defined radius. Outside this region, the gas is stable and remains unperturbed. Earlier images show the same pattern forms for the initial evolution of all the runs, with a circular perturbation moving outwards from the center of the disk which then collapses tangentially to form filaments and knots of gas. The introduction of feedback (bottom two columns) destroys these filaments, smoothing the gas distribution regardless of the type of star formation used. The effect of this smoothing is shown both in the temperature distribution and with the disk's pressure. Hot outflows are seen and cold gas is no longer confined to the dense star forming knots, but rather is blown around the disk. We will see later than this acts not only to change the balance of the phases in the ISM but also to suppress the star formation as knots of gas are destroyed before they can collapse into stars. It also affects the stability in the outer regions of the disk which we will discuss more thoroughly in section~\ref{sec:toomre}. 

The feedback temperature projections also show us the first evidence of the different effects of the C- and D-type star formation routines. The C-type stars cause relatively even outflows (something we will return to in the next section), resulting in a more uniform disk. The D-type, on the other hand, concentrates the outflows, and so we see one side of the disk is much colder than the other side. The position of these outflows changes as the disk evolves, but the separation of the phases is always present. This result is unsurprising if we consider the physics of each stellar type; the D-type stars only form in the densest regions, whereupon they form efficiently and in large numbers. This produces a focusing of stellar material which is transferred to a concentrated injection of energy from supernovae. On the other hand, C-type stars form more uniformly, producing a smoother distribution and hence an even injection of energy. This is shown most clearly in the final column displaying stellar distributions. The C-type stars, forming at much lower densities, extend smoothly out until the threshold radius for gravitational collapse. The D-stars, by contrast, form only in the densest areas of the disk, confining them to the central region where the gas has collapsed both radially and tangentially into dense knots. This result differs somewhat from what was found in TB06, where the lighter disk meant that both C- and D-type star formation algorithms produced stars only in the completely collapsed regions, whereas here we see C-type stars forming in the mildly perturbed areas of the disk (see also section~\ref{sec:toomre}).

The feedback outflows also disturb the pressure in the disk. Non-feedback runs C and D show small-scale variations in the pressure distribution, with the dense, star forming knots of gas being at a higher pressure than the surrounding ISM. The inclusion of feedback, especially in the focused D-type star formation, upsets this, showing the large-scale hot outflows to be over-pressurized with respect to the disk. 


Our second ISM type includes background heating in addition to radiative cooling. Figure~\ref{fig:projHD} shows the projections for runs HC, HD, HCFDBCK and HDFDBCK for the same properties as Figure~\ref{fig:projMW}. Unlike the first ISM type, we see a notable difference in the C and D star formation types in the runs without feedback. The C-type star formation produces large voids that contain hot, low density gas. This shape comes out less clearly in the stellar distribution, but the dense star clusters are confined to the filaments surrounding these holes. The holes have largely vanished in the disks that include feedback, although they all display a porous nature. The same holes are also seen in run C, but only during the first 250\,Myrs of the simulation. After that point, the circular wave speeds up with respect to the HC simulation, and decreases in strength. Exactly why this occurs is not clear. One explanation is that the extra heating in the HC case provides an added pressure that stabilized the circular mode. Alternatively, this is a numerical effect and with improved resolution the porous structure would be retained in the C case as well as the HC simulation. Either way, the presence of holes in the disk is an interesting event and one that has been seen both in our own galaxy and, more dramatically, in the HI map of the LMC. Their presence has traditionally been put down to stellar winds and supernovae explosions evacuating the cool ISM \citep{vanderHulst1996}. However, \citet{Rhode1999} was unable to find evidence of remnant star clusters in the center of the HI holes in the irregular galaxy Holmberg II, suggesting that supernovae were not present there. The issue was also investigated theoretically via two-dimensional simulations performed by \citet{Wada2000} of an LMC-type galaxy which suggested that gravitational and thermal instability alone are enough to create a porous ISM and that these are actually disrupted in the presence of frequent supernovae. This would appear to agree well with our findings that the non-feedback run contained holes, but these largely vanish when feedback is included in the simulation. 

The run HD with D-type star formation also shows some evidence of cavities, but not as strongly as in HC. What is noticeable is that the dense knots of gas are far more evident in this image than in the equivalent run D for ISM~\#1 in Figure~\ref{fig:projMW}. The overall filament structure in both HC and HD is reduced compared to the non-heated simulations, suggesting that the effect of heating is to act against the collapse, increasing the Jeans length to allow only the larger perturbation to form knots. These dense knots, however, extend out further in the heated case than the non-heated case, but the smaller perturbations are smoothed out. We will return to this quantitatively when we consider disk stability in section~\ref{sec:toomre}. 

Feedback again acts to smooth the gas distribution, although less effectively than in the CFDBCK and DFDBCK cases, allowing dense knots of gas to survive, again indicative of them being larger and more tightly bound than in the ISM~\#1 case. We see the same symmetry to the feedback outflows with C-type star formation as with Figure~\ref{fig:projMW} and the focused ejections of the D-type simulations, where the star formation is confined to a much smaller area. 

The pressure distribution in the disk is almost entirely isobaric for the runs without feedback, with only the densest knots in the D-type star formation in HD being at a slightly higher pressure. The heating of the low density background gas also raises its pressure and causes it to be in pressure equilibrium with the disk. Feedback again acts to disrupt this in the disk, most markedly in the D-type star formation case where the hot gas is marked by considerably higher pressures.


Our final ISM type is for the disks with an isothermal ISM at a constant temperature of $10^4$\,K. The projections of these runs are shown in Figure~\ref{fig:projID}. We see the effects of stopping the disk from cooling in the size of the knots of gas, which are visibly larger than in either ISM~\#1 or ISM~\#2 due to the higher Jeans length throughout the whole disk. As a result, the fragmentation is limited to the disk's central region which, especially in the case of the C-type star formation, becomes depleted of gas. The resultant cavity looks similar to that found in HC although it is smaller and corresponds to regions of dense star clusters, rather than a lower density of stars. In this respect, the C-type star formation looks more like the D-type, with the majority of star formation in large clusters in the disk center. Later evolution shows this central cavity growing, leaving a gas deficit void. Overall, the disk takes on a smoother appearance.

The production of larger, if fewer, dense gas knots has a dramatic effect on the stellar density for the C-type star formation run. Analysis of the star clusters (described in \citet{Gill2007}) produced by the large knots of gas reveal masses up to $10^{10}$\,M$_\odot$ with $1/2$ mass radii of order a cell size. These huge dense clumps gravitationally interact with close star particles, accelerating a small fraction of them up to 1000\,kms$^{-1}$ resulting in ejection from the disk in high velocity streams. By contrast, the largest star clusters formed in the C run are a factor of ten less in mass and much more diffuse resulting in a significantly smaller gravitational pull on nearby star particles. In the case of D-type star formation, the star clusters are more tightly bound (since they form only in the densest gas) which makes it harder for star particles to be ejected at high velocities. 

Outflows of this magnitude and speed in the IC case are not observed in real galaxies, and this is a point against using the isothermal ISM model with the C-type star formation algorithm. However, the ejection of star particles by cluster interactions raises an interesting question about star formation in the outer parts of the disk. Observationally, low luminosity stars have been found at large radii \citep{Ferguson2002,Boissier2006}, beyond the point where the disk is traditionally gravitationally stable. How these stars got there is an open question but they may have been produced during satellite interactions, or they may have been thrown there from interactions within the disk. Figure~\ref{fig:projID} suggests that interactions between heavy star clusters could potentially produce this effect. 

The pressure projections show the isothermal model to be the least isobaric of the three ISM types. This is unsurprising when we consider that fixing the temperature forces the pressure to mirror the density distribution. We see here that the gravitationally collapsed structures are over pressurized with respect to the disk and voids of gas have low pressure.

\subsection{The Vertical Profile}
\label{sec:vertical}

\begin{figure} 
\begin{center} 
\includegraphics[width=\columnwidth]{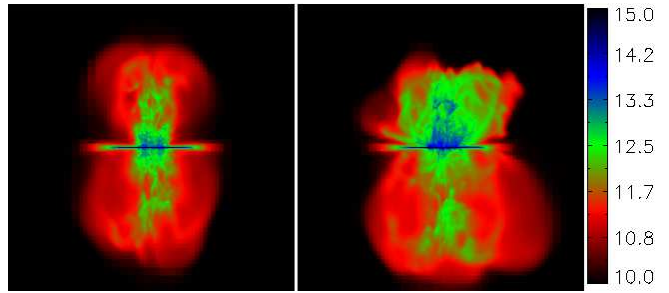}
\caption{Edge-on projections of the baryon density in the CFDBCK and DFDBCK simulations after 142\,Myrs. Images are $\sim 210$\,kpc across. Both simulations include feedback and radiative cooling, but the left-hand image (CFDBCK) has a low density cut-off and low efficiency for star formation whereas the right-hand image shows the disk with a high density cut-off and high star formation efficiency.
\label{fig:VP_Projections}}
\end{center} 
\end{figure}

\begin{figure} 
\begin{center} 
\includegraphics[width=\columnwidth]{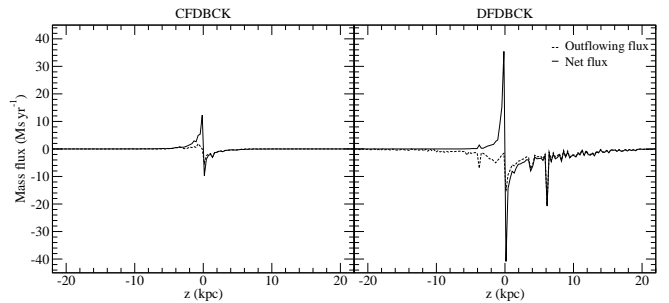}
\caption{Average mass flux out of the disk between t = 283 - 472 Myrs for runs CFDBCK and DFDBCK. Solid line shows the outflow from the disk only while the dashed line shows the net flow above and below the disk.\label{fig:massflux}}
\end{center} 
\end{figure}

The vertical structure of the disk and --- linked with this --- the interaction between the disk and the halo, are thought to play a vital part in the disk's evolution. Indeed, \citet{deAvillez2004} cite the cycle of gas between the disk and the halo as being one of the primary factors in determining the phases of the ISM in their small-box simulations. They find the ability of hot gas to lift off the disk's surface acts as a pressure release valve on the ISM. This is in keeping with the work done by \citet{Norman1989} whose model of the hot gas being removed vertically through galactic chimneys allowed a lower filling factor than in the original picture from \citet{McKee1977}, more in keeping with observations.

Figure~\ref{fig:VP_Projections} shows the vertical projections of the baryonic density in our feedback runs for ISM~\#1, CFDBCK and DFDBCK (the results are similar for HCFDBCK and HDFDBCK in ISM~\#2). The left-hand image shows the simulation with C-type star formation, whereas the right-hand one has D-type stars. Both galaxies are clearly injecting a significant amount of material out of the disk and into the halo, consistent with both \citet{Norman1989} analytical picture of the ISM and \citet{deAvillez2004} small box simulations. We have already seen in Figures~\ref{fig:projMW} and \ref{fig:projHD} that the outflow gas is highly pressurized, in keeping with the release valve idea. This is in contrast with the simulations without feedback, where the gas is confined to the disk's surface. These images are taken shortly after the outflows begin in the disk, at $142$\,Myrs where the difference between the C- and D-type star formation is sharply apparent: D-type star formation not only results in asymmetrical distribution of gas within the disk as seen in section~\ref{sec:images}, but the gas is also ejected unevenly from the disk's surface. This is most likely a result of the density-threshold in the D-type star formation prescription, which can only be met in large clumps and so is naturally highly inhomogeneous. The right hand panel of Figure~\ref{fig:VP_Projections} shows one outflow extending away from the bottom of the disk which has almost reached the image's edge. A second outflow, either just starting up or condensing to fall back in, is shown at the top of the disk. The C-type star formation by contrast produces an even distribution either side of the disk, suggesting gas is being emitted from both sides largely at the same time. 

Despite the symmetrical differences in the gas ejection of the two star formation types, neither run sees the majority of the gas leaving the gravitational pull of the galaxy. Rather, the gas cools above the disk and falls back down in a galactic fountain effect. This can be seen in Figure~\ref{fig:massflux}, which shows both the net mass flux as a function of height from the disk and also the outward-bound only gas. The mass flux is averaged over almost 200\,Myrs, from t = 283-472\,Myrs and over this time range the overall outflow from both C- and D-type star formation is roughly symmetrical above and below the disk. The net mass flux, however, is much smaller than the outflow, showing that the majority of the gas returns to the disk. The size of the outflow is strongly dependent on the star formation type. D-type star formation, with its concentrated clumps, produces a significantly stronger outflow than the C-type, reaching maximum mass fluxes of 30\,M$_\odot$yr$^{-1}$ compared to around 10\,M$_\odot$yr$^{-1}$. In both cases, the outflows are largely restricted to only a few kpc away from the disk. In the C-type case, there is almost nothing outside this region whereas the D-type shows evidence of outflows extending further away from the disk and infalling gas up to heights of 20\,kpc.

\begin{figure} 
\begin{center} 
\includegraphics[width=\columnwidth]{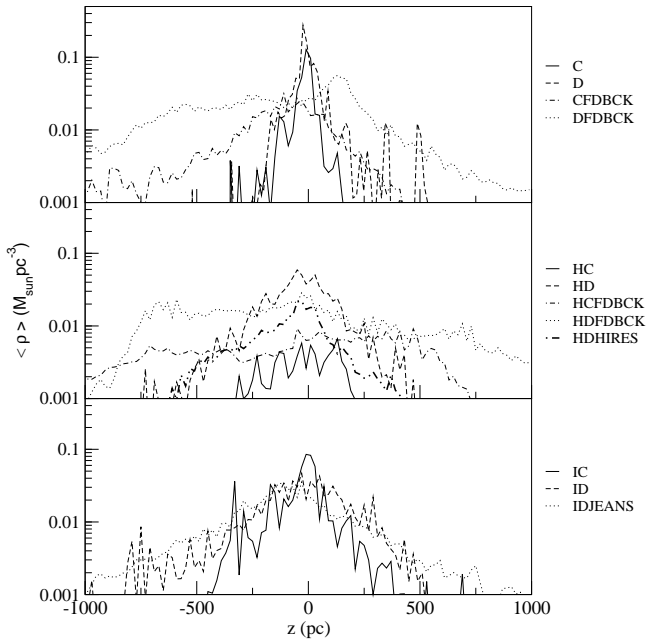}
\caption{The vertical density profile at 472\,Myrs into the evolution of the simulations listed. This is the density profile averaged within two radial scale-heights.  
\label{fig:rhoheight}}
\end{center} 
\end{figure}

\begin{figure} 
\begin{center} 
\includegraphics[width=\columnwidth]{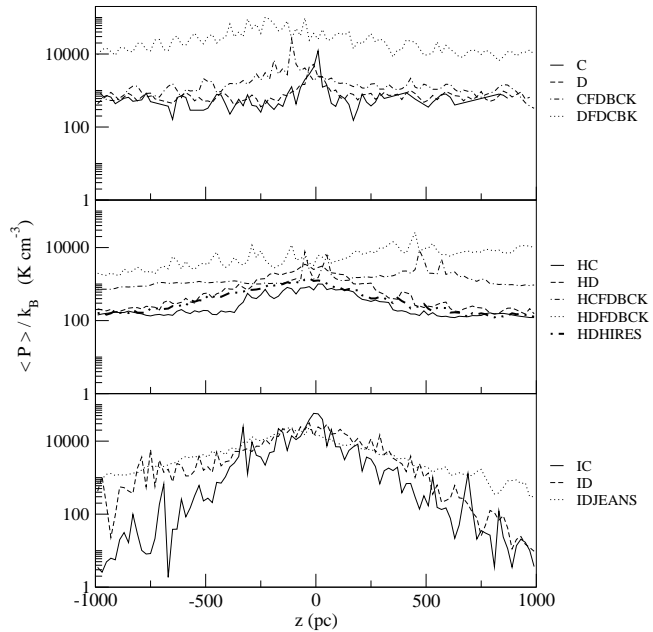}
\caption{The vertical pressure profile at 472\,Myrs into the evolution of the simulations listed. The profile is also averaged within two radial scale-heights.
\label{fig:pheight}}
\end{center} 
\end{figure}

A more quantitative way of studying the vertical distribution in the disk is to examine the one-dimensional profiles. Figure~\ref{fig:rhoheight} shows the variation of baryon density with height after 472\,Myrs for each of the ISM types. The effect of changing the properties of the ISM is very striking in this plot, especially for the runs that do not include feedback. In the top panel, simulations C and D with ISM~\#1 are shown which have a peaked profile with vertical scale height of about $50$\,pc (and which is probably unresolved in the simulation). When background heating is added for HC and HD in ISM~\#2 (middle plot), the disk broadens out, producing a scale height of roughly twice that in ISM~\#1. It is worth noting however, that our background heating term is a constant. A more peaked profile might have been achieved by varying the heating rate with disk height as was done in \citet{Joung2006}. The background heating also affects the two star formation types in different ways; the HD profile is broadened to a greater extent than HC, which becomes flattened in the central most region of the disk. In the isothermal case, the disk is more insensitive to star formation type, although the D-type star formation in ID extends to a higher $z$ than IC. The profile shape and scale height are very similar to HD with $z \sim 100$\,pc. 

The addition of feedback has the same effect on both ISM~\#1 and ISM~\#2, producing a far flatter and broader profile, indicative of the destruction of the central dense gas knots and material being distributed further out from the disk by the outflows. Like ISM type, star formation type has little impact in this case. 

Our higher resolution simulation, HDHIRES, closely follows its lower resolution counterpart, HD. Likewise, the run IDJEANS, specifically designed to resolve where the Truelove criteria for resolution is not met, is almost identical to ID. This is true for all other plots in this paper. 

Over time, the vertical profile remains fairly stable, although runs with C-type star formation decrease in density more quickly than the D-type as the gas is used up in the disk. Feedback acts to reduce this effect and indeed, it suppresses the star formation as we shall see in section~\ref{sec:obs}. Heating the disk also suppresses star formation in the HD run, but has markedly less effect on HC which continues to lose gas. The exception to this are the isothermal runs which are particularly stable, especially ID which shows much less evolution over the course of the simulation.

The vertical pressure profile of the disks are shown in Figure~\ref{fig:pheight}. Most noticeably, runs with ISM~\#1 and ISM~\#2 have largely isobaric profiles as indicated in the projections. The isothermal runs (ISM~\#3) meanwhile have pressure proportional to density and show a clearly peaked distribution. The introduction of feedback in ISM~\#1 and \#2 raises the pressure, corresponding to the high pressure regions we saw in the radial projections in Figures~\ref{fig:projMW} and \ref{fig:projHD}. What is apparent now, however, is that the disk is in greater pressure equilibrium across its height, as gas is ejected from the disk's surface. This shows that the gas ejection seen in Figure~\ref{fig:VP_Projections} is indeed acting as a pressure release system, as described in \citet{Norman1989}.

\section{Phases in the ISM}
\label{sec:ismphases}

\subsection{The Density PDF}

\begin{figure} 
\begin{center} 
\includegraphics[width=\columnwidth]{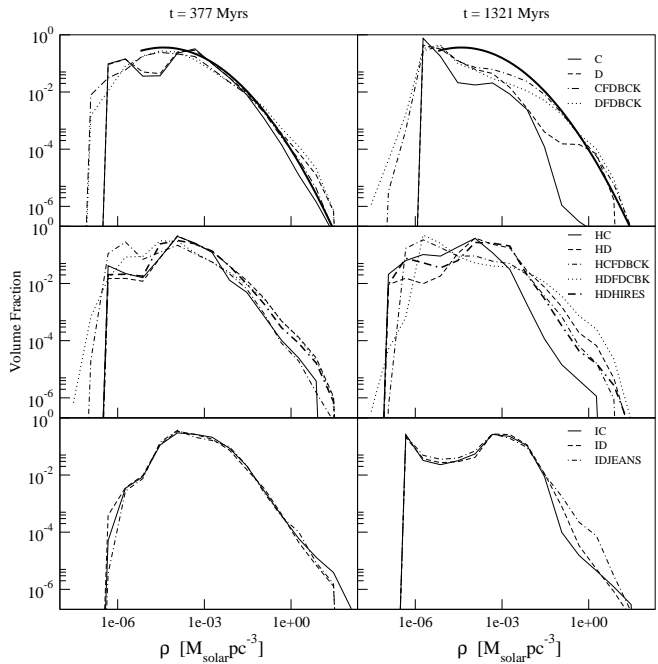}
\caption{Probability distribution function of the volume weighted gas density in the simulated galaxy disks at 377\,Myrs and at 1.32\,Gyrs.  A lognormal distribution is overlayed in the top two panels.\label{fig:volpdfs}}
\end{center} 
\end{figure}

\begin{figure*} 
\begin{center} 
\includegraphics[width=\textwidth]{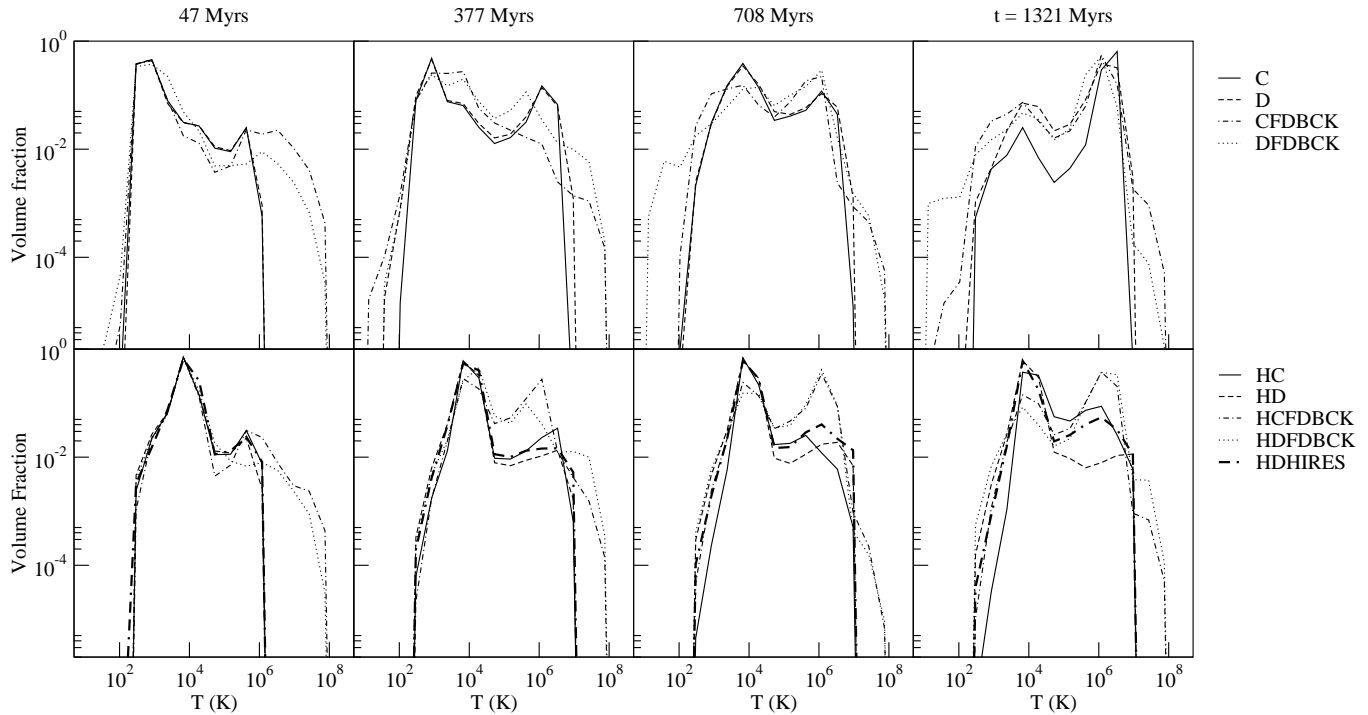}
\caption{PDF of the volume weighted gas temperature for the simulated galaxy disks over the course of the simulation.  
\label{fig:voltemppdfs}}
\end{center} 
\end{figure*}

Section~\ref{sec:discstructure} shows that the different types of ISM models are making a noticeable difference to the appearance of the galaxy disk. This brings us back to our original problem: if the ISM conditions affect the galaxy so much, why do observations show such a simple relation between local gas surface density and star formation properties on large scales?  A way to explain this has been suggested by \citet{Elmegreen2002}, who noted that if the density distribution were everywhere a log-normal distribution, and if star formation occurred above a fixed density, then the Schmidt law would result. This means that despite its complex nature on the very small scales, the fraction of gas dense enough to form stars would always be known from this profile.  \citet{Slyz2005} does point out that this does not completely hold, since no spatial information would be available from such a PDF, resulting in the possibility that the dense regions do not contain enough mass to exceed the critical Jeans limit for collapse. That aside, the presence of a universal density distribution function in our disks could be indicative that such a link between the small and large scales is possible.

Previous simulation work has found evidence that a single profile could be fitted to a range of galaxies, but struggle to agree on the shape. The debate largely centers on whether the PDF is best fitted by a lognormal or a power-law curve at high densities (where star formation will occur). \citet{Scalo1998} puts this discrepancy down to the nature of the gas, finding that an isothermal flow follows a lognormal distribution, while a gas with $\gamma \ne 1$ is better represented by a power-law fit. This view is contradicted in global simulations performed by \citet{Wada2007} and \citet{Wada2001} who do find a lognormal PDF fit without the need for an isothermal gas, although note a steepening at high densities which might be indicative of a power-law tail on the PDF. \citet{Kravtsov2003} also find a lognormal fit for their cosmological simulations of high redshift galaxies as do \citet{Slyz2005} in their 3-D small box simulations, but both note that at the high densities where the fit is applied, the gas is, in fact, nearly isothermal. 

Both the global simulations of \citet{Wada2001} and \citet{Kravtsov2003} show an insensitivity to the input physics, in particular finding that the inclusion of stellar feedback does not affect the PDF shape except at the low density end where \citet{Kravtsov2003} notes feedback produces more low density gas.  Slyz's small box simulations at high resolution show a greater sensitivity to the inclusion of feedback, without which the gas resides predominantly at higher densities.
The normalized volume weighted gas density PDFs for our simulations are shown in Figure~\ref{fig:volpdfs} at two different times; the left column shows the profile at 377\,Myrs, when the disk has settled after the initial burst of star formation, and the right-hand column shows the profile at the end of the simulation, when the majority of the gas has been used up in the disk. The rows separate our three ISM types with the different runs labeled in the key. 

At 377\,Myrs, all three of the ISM types show a similar shape at high density and are reasonably well matched by a lognormal profile. Disks with ISM~\#1 that do not include feedback show a slightly bimodal profile, indicative of a thermal instability \citep{Vazquez2000}. However, the introduction of feedback smoothes this, allowing the disk to follow a lognormal profile even at low densities. Later evolution of all runs shows an increase in the substructure of the disk, with a higher percentage of gas found at lower densities as shown in the right-hand column of Figure~\ref{fig:volpdfs} at 1.3\,Gyrs. The shape of the PDF at this time is similar to that found by \citet{Slyz2005} in their small box simulations with feedback, although our non-feedback simulations in this ISM show the same profile. By this stage, gas depletion is starting to become an issue in the runs without feedback whose high density tail is being eaten away. This is particularly true in the C-type star formation run, C, which shows greater gas loss than the D run. Feedback slows this process considerably and, although does not totally prevent it, it maintains its lognormal fit. The lack of substructure and reduced gas loss suggests that feedback is having a stabilizing influence on the disk. 

The introduction of background heating in ISM~\#2 makes little difference to the profiles at 377\,Myrs, producing slightly more low density gas in the feedback cases. However, the profile shows markedly less evolution than in ISM~\#1 models, maintaining a larger volume at medium densities at the end of the simulation. This is most noticeable in the simulations without feedback which have a significantly higher dense gas content at $t=1321$\,Myrs than their ISM~\#1 counterparts, C and D. The amount of low density gas is less than with runs with ISM~\#1, and slightly greater in the feedback case. The heating term has the strongest effect on the D-type star formation without feedback (HD), reducing gas loss by a factor of 10 from the D case so that it resembles the feedback runs. This results in the lognormal profile being largely maintained in the HCFDBCK, HDFDBCK and HD cases. Our higher resolution run, HDHIRES, shows little difference from its lower resolution counterpart, HD, indicating that these results are not resolution dependent.

The isothermal ISM disks becomes bimodal at around 700\,Myrs, although this is clearly not related to a thermal instability. This peak in the low density gas results from material that is out of pressure equilibrium in areas such the disk center (where star formation has consumed much of the gas) and above the disk where we would expect the temperature to be high. The log-normal fit is slightly less good even at early times here, indicating that a power-law fit might be better. However, this is very hard to judge since like \citet{Wada2001}, we are limited by our resolution. The C- and D-type star formation runs show very little difference in this ISM.

Over the main part of the simulation, our disks show a lognormal profile in the medium and high density gas which is largely insensitive to ISM model, star formation type or the introduction of feedback. This fit stretches over several orders of magnitude in density, in keeping with models by \citet{Wada2007, Wada2001} and \citet{Slyz2005} but contradicting \citet{Scalo1998} statement that the gas must be isothermal to attain a lognormal fit. Substructure in the disk is clearly visible in the lower density gas, indicative of the multiphase nature of the ISM.

At late times, the gas profile fit is eroded by gas depletion in the models where feedback or background heating are not present. The addition of background heating prevents gas depletion for the D-type star formation routine, but has less effect on the C-type algorithm. Our pure isothermal model shows signs of gas depletion but not as marked as in models where cooling is allowed and feedback is not present.  The increase of substructure at later times makes it difficult to determine the best profile fit. The lognormal fit is certainly satisfactory, but a power-law curve might well do as good a job, especially in the case of the isothermal run which seems to show a steeper profile. At our resolution, however, this is still difficult to determine. 

\subsection{Temperature and Pressure Profile Evolution}

In addition to the volume of gas at different densities, the 1D temperature distribution can also provide valuable information about the ISM. Figure~\ref{fig:voltemppdfs} shows the evolution of the temperature over the course of the simulation. Top panel shows the simulations which have ISM~\#1 while the bottom panel shows ISM~\#2. The isothermal model (ISM~\#3) has, of course, only gas at $10^4$ K. 

All the simulations with ISM~\#1 follow a very similar evolution, regardless of star formation type or inclusion of feedback. Initially, the gas quickly cools from the $10^4$\,K starting temperature to the minimum allowed temperature of 300\,K. As the disk fragments and forms stars, the gas is heated by spiral shock waves, infall onto the disk and (in the case of runs which include feedback) supernovae explosions. By 708\,Myrs (third panel), the gas is largely in two temperature pockets: warm gas at $10^4$\,K and hot gas at $10^6$\,K. As star formation continues, the cooler gas is either used up or heated up, leaving only the hotter gas in the disk, which corresponds to the low density peak in Figure~\ref{fig:volpdfs}.

The addition of background heating in ISM~\#2 results in the simulation being significantly more sensitive to stellar feedback, since the background heating has a stronger effect on the non-feedback runs. The added heating stabilizes the warm phase, resulting in the majority of gas remaining at $10^4$\,K over the initial part of the simulation, rather than cooling to 300\,K as in ISM~\#1. In this respect, these disks are most similar to the isothermal case, with the majority of their gas in the stable, warm phase.  The warm phase remains throughout the simulation, but is reduced by the addition of feedback which converts more of the gas into the hot phase. This is a reflection of what we saw in Figure~\ref{fig:volpdfs} where HC and HD maintained a stronger population of medium density gas which we now see is the warm phase, whereas the runs HCFDBCK and HDFDBCK resulted in a great proportion of gas in the low density (and therefore hot) phase.

Again, the difference between the C- and D-type star formation routines is minor, except in the case of HC and HD, where HC contains more hot gas that HD.

\begin{figure} 
\begin{center} 
\includegraphics[width=\columnwidth]{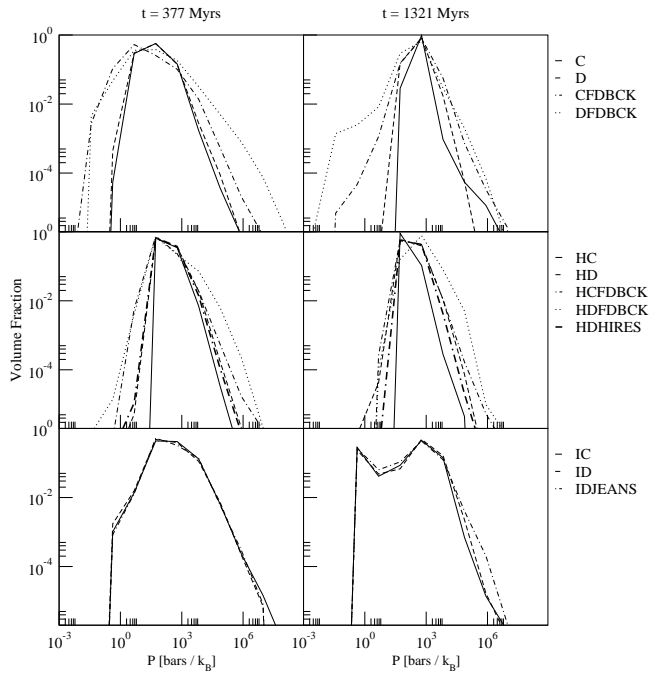}
\caption{PDF of the volume weighted gas pressure for the simulated galaxy disks after 377\,Myrs and end the end of the simulation at 1.32\,Gyrs.
\label{fig:volpresspdfs}}
\end{center} 
\end{figure}

The final 1D study we can make of this type is the volume weighted pressure which is shown in Figure~\ref{fig:volpresspdfs} for the same times as Figure~\ref{fig:volpdfs}. While not completely isobaric, the range of pressures at 377\,Myrs in all the non-isothermal disks is reasonably small. This is especially true for runs with ISM~\#2, where radiative heating raises the temperature of the lowest pressure gas (we will return to this point in more detail in the next section) and least true for ISM~\#3.   Feedback acts to broaden the distribution.  ISMs~\#1 and \#2 show little variation over time, implying the disks are in rough pressure equilibrium. The isothermal case, however, evolves into a bimodal distribution by 1.32\,Gyrs, with a significant quantity of gas at low pressure. This feature is a direct reflection of the density in Figure~\ref{fig:volpdfs} since the temperature cannot change. 

\subsection{The ISM as a Three-phase Medium}

\begin{figure*} 
\begin{center} 
\includegraphics[width=\textwidth]{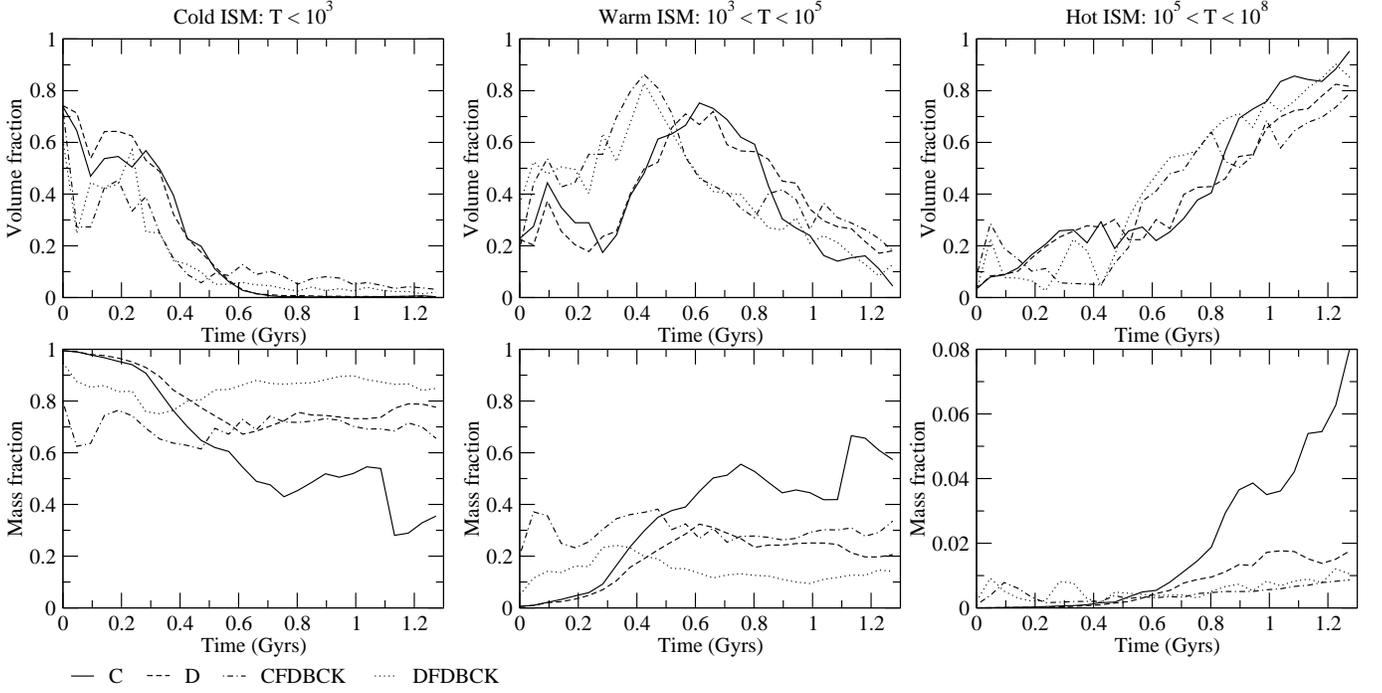}
\caption{Evolution of the cold, warm and hot phases in the disk simulations with ISM~\#1. Here, the cold phase is defined as gas below $10^3$\,K, the warm phase as being between $10^3$\,K and $10^5$\,K and the hot phase as having temperatures over $10^5$\,K.  The mass fraction refers to the gas only. Note different scale for the hot ISM mass fraction.
\label{fig:3phaseMW}}
\end{center} 
\end{figure*}

\begin{figure*} 
\begin{center} 
\includegraphics[width=\textwidth]{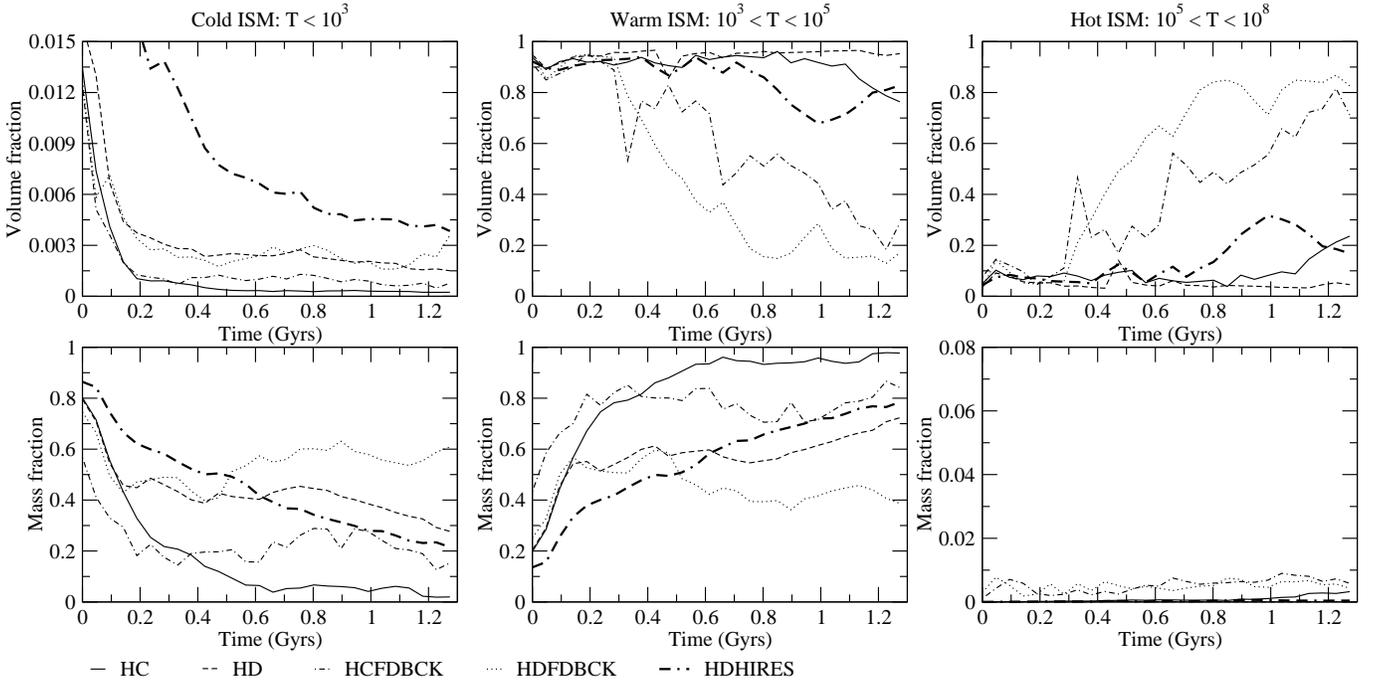}
\caption{Evolution of the cold, warm and hot phases in the disk simulations with ISM~\#2. Temperature ranges the same as in Figure~\ref{fig:3phaseMW}.  Note the different scales on both the cold volume fraction and the hot mass fraction.
\label{fig:3phaseH}}
\end{center} 
\end{figure*}

Taking the idea of the traditional three-phase ISM portrayed by \citet{McKee1977}, we next looked at the evolution of the cold, warm and hot material in the disk over the course of the simulation. Assuming the cold ISM consists of temperatures $< 10^3$\,K, the warm ISM is between $10^3$\,K and $10^5$\,K and the hot ISM is temperatures above $10^5$\,K we look at the evolution of both the volume and mass fractions in the disks.

Figure~\ref{fig:3phaseMW} shows the volume weighted (top) and mass weighted (bottom) evolution for the disks with ISM~\#1. As in Figure~\ref{fig:voltemppdfs} we see the majority of the gas volume initially cools, filling the cold phase. As star formation removes the dense, cold gas, the warm phase begins to dominate. SNe feedback increases the hot phase in the CFDBCK and DFDBCK runs, while the non-feedback cases become depleted still more of gas, allowing the hot phase to dominate which eventually encompasses most of the volume in the disk. The resulting pattern is relatively independent of feedback and star formation type, although the move to the warm phase occurs slightly earlier for runs with feedback since they have the added energy injection.

The gas mass, on the other hand, resides predominantly in the cold phase, sitting in the dense knots of matter that go on to form stars. Despite its large volume, very little mass actually resides in the hot phase (note the change of scale on the hot phase plot abscissa). 

Figure~\ref{fig:3phaseH} shows the same set of plots for ISM~\#2. The difference between the types of run here is much more marked. In all cases however, very little of the gas volume is contained in the cold phase, less than that for the runs in ISM~\#1. This is in agreement with Figure~\ref{fig:voltemppdfs} where we saw a significant proportion of the gas stayed at $10^4$\,K, boosted out of the cold phase by the background heating. For the runs without feedback, the majority of the volume and mass of the gas sits in the warm phase, causing the disk to be closer to isothermal in these conditions. This suggests that star formation is being suppressed in the HD and HC cases, with respect to runs C and D, as the cold star-forming gas is reduced by being moved to the warm phase. We will see in section~\ref{sec:obs} that this is indeed the case. This also causes less of the gas volume to be in the hot phase, since the cool gas has not all been depleted. The feedback runs also start with the bulk of their volume in the warm phase but this is swiftly over-ridden by the stronger effects of the SNe energy injection, heating the gas and moving it into the hot phase.
 
\begin{figure} 
\begin{center} 
\includegraphics[width=\columnwidth]{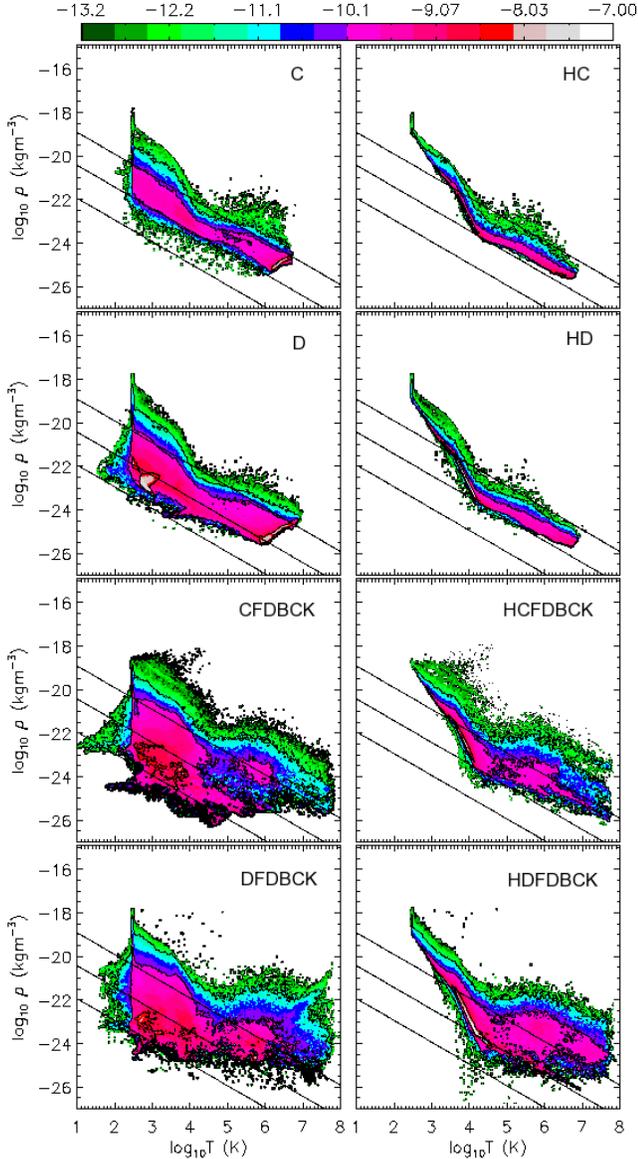}
\caption{Two dimensional contour plots for the volume weighted gas in the runs marked above. Plots are taken at 377\,Myrs.  
\label{fig:volcontours}}
\end{center} 
\end{figure}

A better way to see the phase structure of the ISM is via 2D contour plots of the density versus temperature. Figure~\ref{fig:volcontours} shows the volume weighted contour plots for the disks with ISM~\#1 (left-hand plots) and disks with ISM~\#2 at 377\,Myrs. Straight black lines mark lines of constant pressure (neglecting changes in the mean molecular mass). 

The runs with ISM~\#1 and no feedback (i.e. C and D) show similar behavior: there is rough pressure equilibrium with one or two orders of magnitude scatter along the pressure direction (as seen earlier).  The sharp feature at $\log(T) = 2.5$ is due to our minimum temperature cutoff and the gas that resides in the narrow feature that rises to higher pressure is gravitationally bound.  This is the cold, dense phase.  A peak in the volume distribution can be seen for the high-temperature phase (and less clearly for the warm phase at $T \sim 10^4$ K).  Feedback acts to broaden this distribution, as can be seen in the CFDBCK and DFDBCK runs, making the pressure equilibrium less obvious.

The heated runs (HC and HD) show a very different behavior, with a much tighter relationship between density and temperature.  This is due to the imposition of equilibrium between cooling and heating at low temperatures (below $10^4$ K), which leads to the cold phase having higher pressure.  This tends to force gas into the warm phase and reduces the star formation rate because of the reduced amount of cold, dense gas.  Adding feedback on top of this broadens the distribution somewhat but has a milder effect than in ISM~\#1.

\section{Observational Comparison}
\label{sec:obs}

So far we have seen that the properties of the ISM play a strong role in determining the structure of the disk. We have, however, seen some evidence that a universal PDF might apply to all disks that could allow the observational properties to be largely independent of the gas structure. This section focuses on star formation in our galaxies and compares the results with the main observational relations.

\subsection{Star Formation History}

\begin{figure} 
\begin{center} 
\includegraphics[width=\columnwidth]{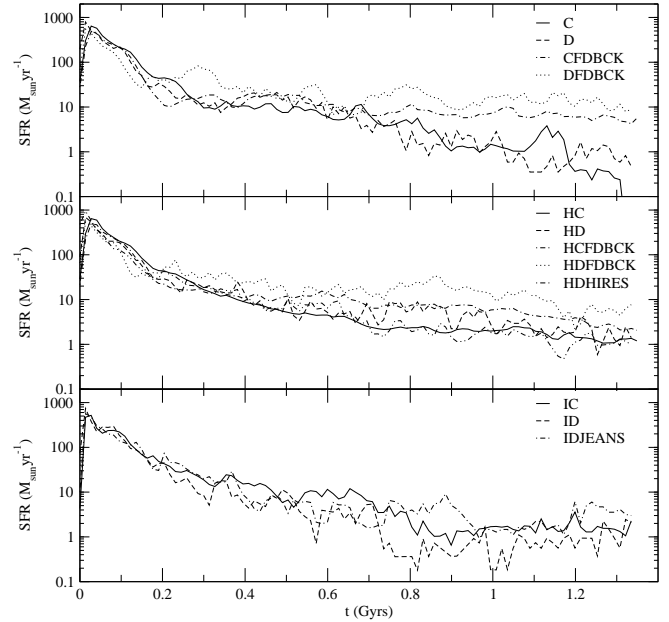}
\caption{Star formation rate over the course of the simulation for the above listed simulations.
\label{fig:SFhistory}}
\end{center} 
\end{figure}

Initially, we turn to the star formation history to examine the rate at which stars are forming over the course of our simulations. This is shown in Figure~\ref{fig:SFhistory} whose three panels depict the evolution of our three ISM types over the course of the simulation. In all cases, the curves follow the same pattern as in TB06; the first 50\,Myrs sees a sudden star burst as the disk becomes gravitationally unstable and starts to collapse. The star formation rate then peaks and falls off roughly exponentially as gas is used up in the disk. 

Considering first the top plot showing disks with ISM~\#1, we see that the runs without feedback have a steadily decreasing star formation rate over the majority of the simulation. In the last 200\,Myrs, the simulation with D-type star formation starts to show signs of reaching a constant value, evening out at roughly 1\,M$_\odot$yr$^{-1}$. The C-type run, C, by contrast, heads to zero star formation and in both these cases, the gas content is significantly depleted as we saw in Figure~\ref{fig:volpdfs}. The time scale for this is significantly less than for present day spirals, as also found in the lighter disk runs of this type in TB06, although of course our initial conditions do not provide an accurate cosmological starting point so this is hard to judge. The simulations that include feedback, however, reach a steady star formation rate at around 600\,Myrs, resulting in a higher star formation rate by 1.3\,Gyrs than for the non-feedback runs. This shows that feedback is causing the star formation to become self-regulated which agrees well with the images of the disks in section~\ref{sec:images} where feedback results the destruction of cold clumps of gas. At the end of the run, the feedback simulations have a constant star formation rate of around 10\,M$_\odot$yr$^{-1}$ which is high compared to the Milky Way, whose star formation rate is roughly 1\,M$_\odot$yr$^{-1}$.  

The effect of adding background heating in ISM~\#2 causes an overall suppression of the star formation rate. In the second panel we can see that the lines are much closer together; the non-feedback runs HC and HD having only a slightly lower star formation rate after 1\,Gyr than the feedback cases. The background heating therefore acts in a similar way to the feedback, suppressing the formation of the cold clumps of matter that form stars. This is again consistent with Figure~\ref{fig:projHD} where the filamentary structure of the gas was notably less than in Figure~\ref{fig:projMW}. Feedback acts to suppress star formation even more, with the lines corresponding to HCFDBCK and HDFDBCK lying above HC and HD, but the overall difference is smaller. Both sets of feedback runs in ISM~\#1 and ISM~\#2 show very similar evolution. At the end of the simulation, all the heated runs have a star formation rate between 1 - 10\,M$_\odot$yr$^{-1}$.

The two isothermal runs are shown in the bottom panel of Figure~\ref{fig:SFhistory}. There is a small distinction between the C- and D-type star formation evolution and, contrary to the other simulations, the C-type star formation is generally higher than the D. This is likely to be due to the C-type stars forming in the larger gas clumps than in the previous ISM models, due to the increased Jeans length, mimicking a behavior closer to the D-type star formation. The star formation rates reach a roughly constant value after roughly 800\,Myrs, giving an end rate of 1\,M$_\odot$yr$^{-1}$, close to what we see for the heated cases, HC and HD, and in good agreement with the Milky Way. 

\subsection{The Star Formation Cut-off}
\label{sec:toomre}

\begin{table}
\caption{Star formation cut-off (radius which includes 99 \% of sf) and Toomre Q value at that point.}
\begin{tabular}{cccc}
& Cut-off radius (kpc) & Toomre $Q$ & Obs. $Q$\\ 
\hline
C      & 13.3 & 0.40 & 0.82\\
D      & 11.4 & 0.26 & 0.54 \\
CFDBCK & 10.7 & 0.30 & 0.47 \\
DFDBCK & 12.4 & 0.28 & 0.58 \\
HC      & 17.8 & 6.41 & 3.49 \\
HD      & 13.8 & 0.95 & 0.81 \\
HCFDBCK & 13.6 & 1.02 & 0.86\\
HDFDBCK & 15.0 & 1.45 & 1.07\\
HDHIRES & 16.1 & 1.78 & 1.20 \\
IC      & 12.2 & 1.83 & 0.72 \\
ID      & 12.4 & 1.91 & 0.76 \\
IDJEANS & 12.9 & 1.87 & 0.74 \\
\end{tabular}
\label{table:toomre}
\end{table}

The gravitational collapse of gas into cold knots of matter is the fundamental driving force for star formation in our galaxies (this is discussed more thoroughly in TB06 and references therein). Previous studies, both theoretical and observational, suggest that at a given radius the density of the gas drops below a given critical value and the disk becomes stable, preventing star formation from occurring beyond this point. Toomre \citep{Toomre1964} initially defined the location of this radius in terms of a stability parameter, $Q$, given by $Q=\kappa c_s/\pi G \Sigma_g$, where $\kappa$ is the usual epicyclic frequency, $c_s$ is the thermal sound speed as measured in the disk and $\Sigma_g$ is the gas surface density. Toomre's calculations considered axisymmetric perturbations in a single phase two-dimensional disk which was found to became gravitationally unstable (and therefore able to form stars) when $Q < 1$. \citet{Goldreich1965} re-calculated this value for a three-dimensional disk to $Q < 0.67$. Observational results have also observed this star formation cut-off, with \citet{Kennicutt1989} measuring the star formation threshold in spiral galaxies to correspond to a $Q$ value of $1.5$. Observations did not allow measurement of the actual thermal sound speed of the gas, which was replaced by a velocity dispersion of $6$\,kms$^{-1}$.   

In our disks, we measured both the radial cut-off point for star formation and the Toomre $Q$ parameter for the disk's gas at that radius. We also calculated an observed $Q$ value, where we adopt Kennicutt's value of $6$\,kms$^{-1}$ instead of the thermal sound speed. It is worth noting, however, that these calculations were not entirely simple. The greater gas mass and smaller star particle size results in a significant gravitational scatter of stars near the edge of the stellar disk than we had in TB06, so the exact edge of star formation is hard to judge. (This is of course true in observational result too, a point we will return to in the next section). We therefore take the radius where 99\,\% of the star particles are enclosed as the stellar cut-off. The measuring of $Q$ in a multiphase disk is an even harder task to perform accurately. Firstly, in averaging over an annulus at a given radius, you include a wide range of temperatures and densities which produce an average $Q$, not necessary the $Q$ value at the star formation sites. Secondly, Toomre's original calculations assumed linear perturbations which break down at the point of star formation. Ideally, therefore, you want to measure the $Q$ value where the disk has become gravitationally unstable, but not yet formed stars. In TB06, such a point existed, but with our heavier disk and smaller star particles the disk begins to form stars earlier on. We therefore measure $Q$ at the same time as in TB06, approximately 50\,Myrs after the start of the simulation and note that while the disk has fragmented, stars have already started to form. 

What is surprising is that we do measure only a small range of values for both the cut-off radius and $Q$ as shown in Table~\ref{table:toomre}. In all cases, the star formation ends at a comparable radius to the Milky Way, which is estimated to be around 15\,kpc. Within the range we do see some patterns. For the non-feedback cases, the C-type star formation simulations form stars out to a radius several kiloparsecs greater than in the D-type, something we saw visually in Figures~\ref{fig:projMW} and \ref{fig:projHD}. The reason for this is that the D-type star formation's higher density threshold, confining stars to regions where the gas has collapsed to form the dense knots of matter. Our C-type star formation however, can occur at much lower densities, when the gas has not fully collapsed. The exception to this is the isothermal gas, where, as previously mentioned, the larger Jeans length confines the C-type stars to the same region at the D-type. The addition of feedback has a different effect on the two star formation types. For the C-type, it reduces the threshold radius whereas for the D-type, the radius for star formation increases. This difference is caused by the nature of the resulting outflows from the feedback. As we saw in Figure~\ref{fig:VP_Projections}, feedback from C-type stars produces a smooth, continuous outflow that disrupts the dense gas and prevents stars from forming. Feedback from the D-type stars however, comes in energetic bursts which, as we saw in Figure~\ref{fig:massflux}, extends the gas outflows to much higher radii. These outflows of gas can then cool and condense, fragmenting beyond the old instability threshold to push star formation out to higher radii. 

A slightly surprising result is that the heated disks have higher threshold radii than the non-heated cases. Since the heating has a stabilizing effect on the disk (as seen in Figure~\ref{fig:SFhistory}) we would expect it to quench star formation in the less dense outer disk. We actually see in Figure~\ref{fig:projMW} that the instabilities do extend out further in the DC and DD case than the HDC and HDD, but the star forming knots stop at a lower radius. This appears to be the result of the dissipation of the circular wave that we noted occurred in DC in section~\ref{sec:images}. Although, as mentioned, the exact cause of this is unclear.

If we look at what $Q$ parameter this cut-off corresponds to, we find values around 1. Given that $Q$ itself varies over several orders of magnitude, the results are very uniform, with neither ISM conditions, feedback or star formation type having a large effect. The only value that appears out of place is for run HDC with $Q=3.49$. Given the range of $Q$ over the disk, this value is still in keeping with the other simulations although its higher value may be due to problems accurately measuring the star formation cut-off.

Coupled with the lower star formation cut-off, disks without heating in ISM~\#1 have lower $Q$ values than the ISM~\#2 and ISM~\#3 disks. Feedback and star formation type have little effect on these values, which are all below 1. ISM~\#2 shows the most sensitivity to stellar conditions, with the C-type star formation in particular being more sensitive to the introduction of feedback than the D-type. The isothermal ISM~\#3 simulations show almost no variation with star formation type. For comparison, the third column in table~\ref{table:toomre} shows the observationally calculated $Q$ parameter. The values are reassuringly similar to the first calculation of $Q$, although we can see that for ISM~\#1, the $6$\,kms$^{-1}$ is an overestimate of the thermal sound speed, since it raises the value of $Q$, whereas in the other cases it is an underestimate. 

\subsection{The Global Star Formation Relation}

\begin{figure} 
\begin{center} 
\includegraphics[width=\columnwidth]{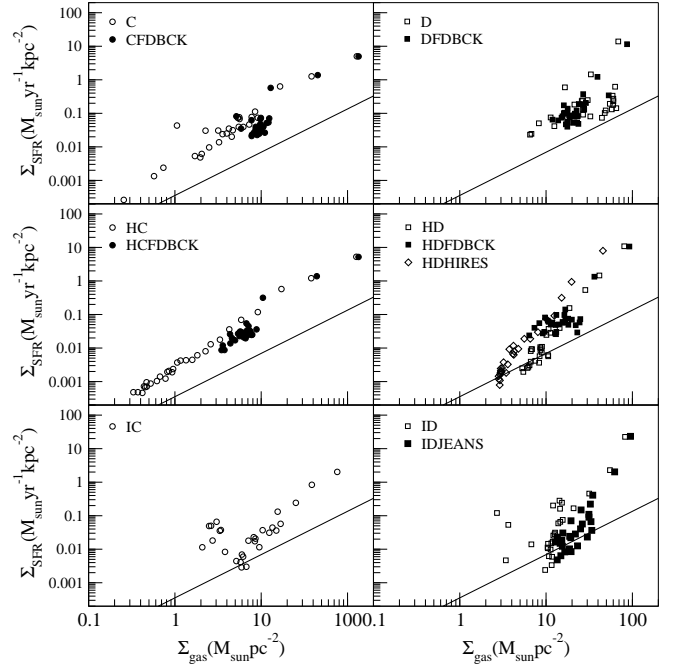}
\caption{Surface star formation rate versus surface gas density averaged over the whole disk for different times during the simulations. This is the Global Schmidt law and the solid line shows the best fit from observations \citep{Kennicutt1989}.
\label{fig:globalschmidt}}
\end{center} 
\end{figure}

Kennicutt's observations of disk galaxies show that the surface star formation rate is linked, both on a global and local scale, to the surface gas density via $\Sigma_{\rm SFR} \propto \Sigma_{\rm gas}^{1.5}$. In Figure~\ref{fig:globalschmidt} we show Kennicutt's observational relation (solid line) and the results from our different simulations for the global version of this law (Global Schmidt law), which shows the average surface star formation rate and the average gas surface density over the whole disk for different (evenly spaced) times over the $1.4$\,Gyrs of the simulation. 

Overall, the majority of the simulations reproduce the $1.5$ gradient well, although all over estimate the star formation rate. The possible exception to this is the isothermal run with D-type star formation, IDD, whose points contain a large amount of scatter and a steep decline at a gas density of $10$\,M$_\odot$pc$^{-2}$.

The over-estimate in the SFR was discussed in TB06 and is likely due, at least in part, to our inability to include all the physics at the resolution of the giant molecular clouds. Without the added destructive effects of ionizing radiation and stellar winds, these clouds lifetimes will be prolonged, allowing an unphysically high proportion of their gas to be converted into stars. The other possibility is that neither of our star formation mechanisms is very accurate. To a certain extent this is inevitable since we do not resolve the GMCs, but recent work in this area \citep{Krumholz2006} suggest that while star formation occurs in the dense molecular clouds, on scales larger than a few parsecs the star formation efficiency is much lower than the 50\,\% suggested by \citet{Lada2003}, possible down to 2\,\%. This reduced efficiency will be explored in future papers. 

As in TB06, we find the addition of feedback does act to reduce the star formation rate, confirming that we are seeing at least the beginning of self-regulation. This is most evident in the simulations with C-type star formation: C, CFDBCK, HC and HCFDBCK, all of which lie close together on the graph. Simulations with D-type star formation also follow the observed gradient well, especially in the non-heated simulations, D and DFDBCK. This is important since, as discussed in TB06, the C-type star formation has a Schmidt-like behavior built into it, implying that we are simply getting out what we put in. However, this is not true for the D-type algorithm, which has a simple density cutoff and is otherwise proportional to the density.

As hinted at with the PDFs in Figure~\ref{fig:volpdfs}, the background heating in ISM~\#2 has a stronger effect on the HD simulation than on HC, as the higher density regions required for this type of star formation are disrupted by the increase in temperature. This results in the star formation at low densities in HD to be significantly reduced from D. At these low densities, the results from HD now follow the observations very closely, but at higher densities, the star formation rises above what is expected. This suggests that the background heating has less effect at the higher densities, where the gas is collapsing despite the extra resistance.  Introducing feedback reduces this effect, still producing an over-estimate of the star formation, but the results are now consistent with HCFDBCK, CFDBCK and DFDBCK, implying that feedback is the dominant effect here. 

The isothermal runs show the poorest agreement with observation here. The C-type star formation, IDC, produces a gradient slightly steeper than 1.5 for gas densities greater than around 6\,M$_\odot$pc$^{-2}$ but has a large scatter below that. 

\subsection{The Local Star Formation Relation}

\begin{figure} 
\begin{center} 
\includegraphics[width=\columnwidth]{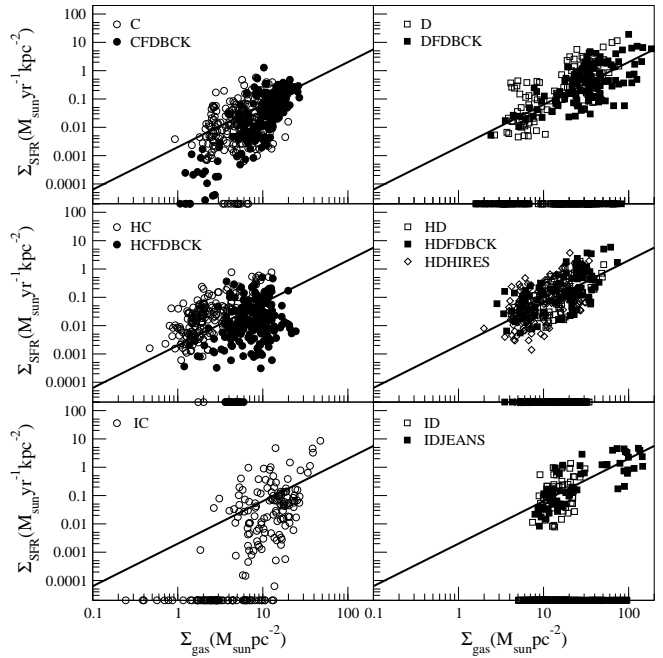}
\caption{The local Schmidt law shows the variation of surface star formation rates with gas surface density at different radii of the disk averaged over times between t = 283 - 472 Myrs. The solid line is a curve with slope 1.5, as observations indicate (but with arbitrary normalization).\label{fig:localschmidt}}
\end{center} 
\end{figure}

The Schmidt law can also be looked at on a local scale where the values are plotted as a function of disk radii rather than time. Figure~\ref{fig:localschmidt} shows the results for each of the simulations where each point represents a different radii in the disk. The solid line has a gradient of 1.5, in agreement with the observations, but with an arbitrary normalization. 

In the top panel where we show the simulations with ISM~\#1, we can see that runs with D-type star formation, D and DFDBCK, represent the observed gradient of the correlation between the surface gas density and star formation rate extremely well. Runs with C-type star formation do a reasonable job at high densities, showing a larger degree of scatter than the D-type star formation runs due to clumps being split by energy injection and gravitational interactions and reforming (something that is not observed in the D-type run due to the higher binding energy of the dense knots). At lower densities, the feedback run, CFDBCK, does a noticeable poorer job at reproducing the Schmidt law, showing a sharp decrease in the star formation rate below gas densities of around 2\,M$_\odot$pc$^{-2}$, a feature that is not seen in the D-type runs which are not affected by the introduction of feedback. This feature is not seen when we introduce background heating in ISM~\#2 (middle panel) where all simulations follow the observations very well, although there is again noticeably more scatter in the C-type run with feedback than in either the run without feedback or the D-type simulations, HD and HDFDBCK.

The isothermal run with D-type star formation follows the same pattern, and indeed there is no difference between D, HD and ID simulations, showing that the D-type star formation is much less influenced by its interstellar environment. The isothermal run with C-stars, however, does not follow the observations well, producing a gradient that is much steeper than observed thereby underestimating the star formation at low densities and over estimating it at high densities. Overall, the range in gas densities is much less than in other C-type runs, agreeing with the results in section~\ref{sec:images} that star formation is confined to a small section of the disk. 

In both the global and local Schmidt law plots, the C-type star formation extends to much lower densities than the D-type. Although this is not surprising, since its density threshold is lower, it is interesting to compare this to observations. As mentioned in the previous section, \citet{Kennicutt1989} observed a sharp cut-off in the star formation in disk galaxies below a critical density. More recent observations done using the UV data in GALAX \citep{Boissier2006} suggest that this star formation cut-off might be a result of the observational technique, rather than the existence of a critical density. \citet{Boissier2006} looked at 46 spiral galaxies in the UV to establish where star formation ended. Previous work had been performed by examining the H$\alpha$ emission from galaxies, a technique, this group argues, that makes it very difficult to observe low levels of star formation. With UV spectra, they found evidence of star formation beyond the position of the previously measured cut-off radius suggesting that the critical density value is either lower than originally measured or possibly does not actually exist. If this is the case, then our C-type star formation is preferable in this result, since it allows star formation to occur at much lower densities in agreement with this new result. 

\section{Discussion}

What role does the interstellar medium play in determining the star formation properties of galaxies? This question has two important consequences. The first concerns the nature of galaxy evolution, asking whether the interstellar environment can result in dramatically different structure and star formation properties. The second consequence applies to our ability to model galaxy formation realistically or whether parsec resolution is needed to achieve accurate results. 

In section~\ref{sec:discstructure} we visually examined projections of the galaxy disks looking at their density, temperature and pressure distributions and their star formation. Clear differences were seen between our disks, with ISM type, star formation algorithm and feedback all playing significant roles. The growth of perturbations in the disk was dominated by the conditions in the ISM. In our first ISM type, which contained only cooling, we saw the disk fragment through the production of a outgoing circular wave which collapsed tangentially to form the network of filaments and dense knots of gas we saw in Figure~\ref{fig:projMW}. The addition of background heating in ISM~\#2, suppressed the formation of the weaker filaments but increased the size of the densest knots which extended further out into the disk. In our third, isothermal, ISM type, the filamentary structure was also suppressed and the size of the dense knots increased still more, confining their formation to the central region of the disk. 

These differences made a difference when star formation started to occur. In our C-type (cosmological) algorithm, stars formed over a large fraction of the disk's surface, although the largest clusters were confined to the densest knots of gas. In the heated case, the heavier filaments formed edges around voids of low density, hot gas that had occurred through gravitational and thermal instability alone. Such voids are reminiscent of porous ISMs seen in the HI maps of some dwarf galaxies. In ISM~\#1, further fragmentation had prevented this from occurring. The isothermal disk confined the stars to the central disk region where the large knots formed and the resulting heavier clusters scattered the smaller ones out of the disk. The extent of this scattering was extreme, and care should be taken if an isothermal model is used with this star formation recipe to avoid unphysical results. The effect in ISM~\#3 of confining the star formation to the central region meant that there was little difference between the C-type algorithm and the D-type star formation algorithm where we restricted the formation of stars to the densest clouds. This was investigated quantitatively in section~\ref{sec:toomre} where we measured the star formation cut-off radius and compared it to the Toomre stability criterion $Q$ at that point. In the ISM~\#3 case, the cut-off radius was almost identical in both the C- and D-type star formation routines whereas in ISM~\#1 and ISM~\#2, the C-type stars extended further into the disk. The $Q$ stability criteria itself was found to be around 1, in good agreement with both analytical and observational results. 

This changed again when we included feedback. In both the ISM~\#1 and ISM~\#2 cases, feedback destroyed the dense knots of gas, suppressing the star formation as we saw in Figure~\ref{fig:SFhistory}. Its addition largely wipes out the structural differences between the two ISM types and their profiles, both in the projections and the vertical profiles in Figures~\ref{fig:rhoheight} and \ref{fig:pheight} follow similar shapes. Feedback also causes gas to be ejected from the disk's surface in a galactic fountain, something the isothermal disk is incapable of replicating. The different star formation algorithms have a much larger effect on the feedback than ISM type. The D-type algorithm, confining the star formation to the densest knots, focuses the energy injection from the feedback, causing the outflows to be bursts in different areas of the disk. The C-type, by contrast, acts equally across the disk and at a more continuous level. This has two effects on the disk. The first was seen in Figure~\ref{fig:VP_Projections} where the outflows from the disk's surface are uneven. The second is seen in the star formation cut-off, where the feedback in the D-type case triggers fragmentation of the gas in the stable regions of the disk, extending the star formation threshold. 
 
Section~\ref{sec:ismphases} further examines the ISM by looking at the evolution of its density and temperature over time. We see that background heating stabilizes a significant proportion of the gas in the warm phase in Figure~\ref{fig:3phaseMW}, bringing it nearer to the isothermal state as can be seen in the 2D contour plots in Figure~\ref{fig:volcontours}. This could explain the success of models using an isothermal ISM; it is a gross simplification of the real system, but if a single phase becomes stabilized in the disk, an isothermal gas can be a good approximation. The addition of feedback, however, dramatically increases the gas volume in the hot phase causing the heated and non-heated disks to show a wide variety of continuous phases. Both ISM~\#1 and ISM~\#2 produce disks that are largely isobaric, in keeping with analytical expectations. The addition of feedback acts as a pressure value to eject gas from the disk, ensuring the disk remains isobaric across its height. The isothermal disk, by contrast, has strong pressure variability since the fixed temperature means that the pressure has to reflect the density distribution. 

In section~\ref{sec:obs} we turned to look at the observable properties of our disks, including star formation history, cut-off and the relation between gas surface density and surface star formation rate (Schmidt laws). No one model produces these results significantly better than any other. A result that implies, despite structural differences, that global star formation is not strongly dependent on the interstellar environment. The disks overall produce the observed properties reasonably well, having a star formation threshold of between 11-18\,kpc in agreement with estimations for the Milky Way and producing the Schmidt observed gradient of 1.5. The slight exception to this were the isothermal cases which reproduced the observational result poorly at low densities. Of the two star formation algorithms, each has its own strengths and weaknesses. The C-type algorithm allows the extension of the Schmidt law to much lower densities, where recent observational evidence suggests there is star formation. However, the small stellar clusters gravitationally interact to be scattered in the disk, making it more difficult to determine the cut-off density for star formation and increasing the scatter in the local Schmidt relation. The D-type algorithm, meanwhile, show a tight correlation in the local Schmidt relation but appears to reproduce the gradient of the global relation less well. However, it is also the least affected by ISM type, since it considers only the densest gas structures. Possibly a hybrid of these two schemes would combine their strengths to produce the best results. 

\begin{figure} 
\begin{center} 
\includegraphics[width=.35\textwidth]{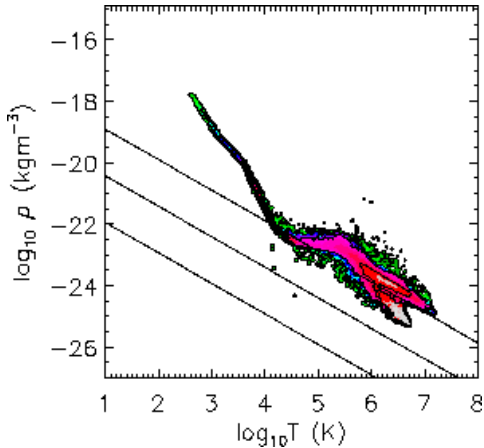}
\caption{Two dimensional contour plots for the volume weighted gas in the run with increased heating source. Plots are taken at 377\,Myrs.  
\label{fig:hd2_volcontours}}
\end{center} 
\end{figure}

As mentioned in section~\ref{sec:enzo}, we performed an additional run with a higher background heating source. This run was identical to run HD, but with the background heating increased by a factor of 30. The main difference the additional heating had on the disk was to increase the pressure of the ISM.  This occurred because the enhanced heating rate moved the gas in thermal equilibrium to higher temperature at fixed density, pushing the contours shown in Figure~\ref{fig:volcontours} to the right and up.  This can be seen in Figure~\ref{fig:hd2_volcontours}. The effect this has on the star formation, however, was fairly minimal. There was a slightly increased star formation rate at early times which led to a greater gas depletion near the end of the simulation. This resulted in the global Schmidt law following a relation identical to that seen in the HD run, but at a slightly raised star formation rate.

Ultimately, HI images of the whirlpool galaxy and LMC tell us that the ISM plays an important role in the galaxy's evolution and it must be born in mind that while we are able to resolve a multiphase structure in our ISM, we are no yet at the detail of the small box simulations. However, the uniformity of our results for the star formation history and Schmidt laws suggest that the exact details of the gas may be simplified and still achieve a correct star formation production. 

\section{Conclusions}

We performed high resolution simulations of global disk galaxies with three different interstellar medium properties: our first model allowed radiative cooling of gas, the second model allowed radiative cooling and included a background heating term while the third model held the gas at a constant temperature. For each of these ISM models, we considered two different star formation mechanisms; a `C' type that allowed low density formation of stars at a correspondingly low efficiency and a `D' type that only allowed star formation in the densest clumps, but with a high efficiency. We also investigated the addition of feedback from type II supernovae for non-isothermal disks. Overall we found:

\begin{enumerate}
\item The structure of the disk is strongly affected by the ISM environment. Background heating suppresses small-scale perturbations and increases the size of the star forming knots. This additionally helps to regulate star formation and reduce gas depletion in the disk. The isothermal equation of state increases the Jeans' length in the disk and leads to the formation of much larger clumps (probably unphysically large) which results in extreme gravitational scattering of nearby star particles.  It also confines star formation to the central region, regardless of the critical density specified in the star formation routine.  This leads to gas depletion in the disk center. 

\item The addition of feedback destroys star-forming clumps and causes gas to be ejected off the disk's surface. The nature of the outflows is dependent on the star formation algorithm with the low threshold density, C-type routine producing a more continuous flow compared to the D-type strong intermediate bursts. Both cases result in star formation being suppressed in the disk as dense knots of gas are destroyed, but the feedback in the D-type case results in triggered star formation in the outer regions of the disk. 

\item The structure of the ISM in both ISM~\#1 and ISM~\#2 show a continuous range of densities and temperatures that are not well represented by a single phase model. The addition of heating reduces the range of values found in the non-feedback case, in particular stabilizing the warm phase at the same temperature as the isothermal ISM. However, the introduction of feedback significantly increases the range of densities and temperatures, almost eliminating the signature of background heating.

\item The PDFs for all disks are well represented by a lognormal curve over several orders of magnitude. This is largely insensitive to the introduction of stellar feedback or ISM environment. At later times (and for the isothermal run at earlier times) a power-law fit would also be possible and it is hard to tell, at current resolutions, which of these two fits would work best. 

\item The star formation in the disks are shown to cut-off at a radius comparable to the Milky Way when the Toomre $Q$ parameter is around 1. Exact measurements are difficult to achieve in a multiphase medium. 

\item All simulations reproduce the slope of the observed relation between star formation and gas surface density well on both the global and local scale. The possible exception is the isothermal run with C-type star formation which shows a steeper drop off in star formation rate on local scales. There is some evidence that D-type star formation does not produce the gradient as well at C-type on global scales and that its low density cut-off may be higher than recent observations. 

\end{enumerate}
Overall, we conclude that the interstellar medium is a multiphase environment strongly affected by background heating, feedback and cooling and plays a significant part in the disk's structural evolution. It appears, however, from these preliminary calculations that its exact structure can be simplified and still achieve the correct star formation properties for disk galaxies. The exception to this is possibly the isothermal disk which did not produce a realistic multi-phase model, lead to very large clump formation, and can never produce a galactic fountain effect. \\*[0.5cm]

\acknowledgements
EJT and GLB would like to thank Adrianne Slyz for helpful comments and suggestions and the referee for helping to improve the presentation of the paper. EJT acknowledges support from a Theoretical Astrophysics Postdoctoral Fellowship from Dept. of Astronomy/CLAS, University of Florida. Both authors acknowledge the National Center for Supercomputing Applications and the University of Florida High-Performance Computing Center for providing computational resources and support and also support from NSF grants AST-05-07161, AST-05-47823, and AST-06-06959 that have contributed to the research results reported within this paper.

\end{document}